\documentclass[pra,twocolumn,showpacs,superscriptaddress,floatfix, reprint]{revtex4-1}

	\usepackage{graphicx}				
	\usepackage{float}					
	\usepackage{color}
	\usepackage{dcolumn}				
	\usepackage[cp1252]{inputenc}		

	\usepackage{bm}						

	\usepackage{physics}				

	\usepackage{xcolor}					

\begin{document}


\title{A shape preserving atomic pulse amplifier}

\author{Nilamoni Daloi}%
\email{nilamoni@iitg.ac.in}
\author{Tarak Nath Dey}
\email{tarak.dey@iitg.ac.in}
\affiliation{Department of Physics, Indian Institute of Technology Guwahati, Guwahati 781039, Assam, India}

\begin{abstract}

Propagation of a weak probe pulse through a $\Lambda$ system in a resonant gain configuration is investigated.  We employ the control field intensity that permits the amplification of probe pulse during propagation, without instability at two photon resonance. Posterior to amplification, a broadened probe pulse is obtained, which retains its initial pulse shape and travels at the speed of light in vacuum, without experiencing any delay, absorption and dispersion. The salient feature of this technique lies in the fact that in addition to preserving the initial pulse shape, it also ensures stable pulse propagation after amplification. It also works for arbitrary probe pulse shapes.

\end{abstract}

	\date{\today}

	\maketitle



\section{\label{sec:level1}Introduction}
Light pulse generation, reshaping and shape preserving propagation has received considerable attention in multilevel atomic systems due to its potential applications in optical communication and information sciences \cite{RN16567,Boyd1,lpr1}. Population inversion \cite{Olga_1990} and atomic coherences are two main constituents for the pulse generation and reshaping \cite{Lukin_2000}. Coherently controlled light-matter interaction can produce desired atomic coherences and population among the various level systems. A variety of techniques based on stimulated Raman adiabatic processes (STIRAP)\cite{stirap_1}, electromagnetically induced transparency (EIT)\cite{eit_1}, coherent population trapping (CPT)\cite{cpt1}, and saturated absorption \cite{sat1} has been adopted to control the dynamics of population and coherences effectively. The induced coherence among the lower level states of a molecular system \cite{PhysRevLett.81.2894} or atomic system \cite{PhysRevA.94.053851} are the central issue of producing pulse radiations. Various techniques have been proposed for the generation of optical pulses \cite{pulse_generation1,pulse-generation2}.\\ 
Self induced transparency (SIT) is a prominent example for shape preserving ultrashort optical pulses propagating through a resonant medium \cite{McCall1,McCall2}. The physical explanation of SIT comes from re-radiation of atoms from the excited state of the two level system in the presence of pulse power which is beyond some critical value \cite{McCall2}. This lossless shape preserving propagation can be extended to  two pulses in an otherwise opaque three level medium which are referred to as simultons \cite{PhysRevA.24.2567}.  Remarkably, the two pulses may copropagate as  complementary pulse shapes in a three level $\Lambda$-system which is dictated by the input envelope shape \cite{PhysRevLett.73.3183}. 
An extensive study on co-existence of stable propagation of $\sech$ and $\tanh$ pulse envelopes have been investigated  in three- and multilevel systems \cite{Eberly_1995,PhysRevA.58.R805,PhysRevA.57.4643,PhysRevA.60.4187,agarwal1}. 
The three-level $\Lambda$ configuration also offers the temporal cloning of an arbitrary shaped strong pulse envelope on to a weak field \cite{cloning1}. Recently the predictions of simultons \cite{PhysRevA.24.2567} have been experimental verified in a V-type thermal atomic system referred to as a quasisimulton \cite{simultons2}. In these studies, radiative and non-radiative decay is not taken into consideration  due to the short duration of the pulse.  However the influence of various atomic relaxation processes plays a crucial role on spectral bandwidth of the generated pulse and the propagation velocity \cite{MATSKO2001191}. The weak probe pulse along with strong control field propagate as matched pulses through an absorptive medium \cite{matched1}. All of these studies mostly requires optical pulses to be of some very specific input forms or definite energy in order to demonstrate shape preserving propagation throughout the length of the medium \cite{Rostovtsev_2011}.\\
Media which exhibit huge absorption, limits any realistic application based on pulse generation and shape preserving propagation.  The absorptive medium accompanied by its inherent pulse broadening, distortion as well as suppression of output transmission, practically stops its usage \cite{Tanaka_2003}.  Hence coherent manipulation of absorption of the medium is essential for arbitrary shaped optical-pulse propagation with controllable width and gain \cite{Wang_2000,PhysRevA.64.053809}. 
Much attention has been paid to make an opaque medium transparent for supporting propagation of pulses without changing their initial profiles \cite{Eberly_2008,Gordon_2018,Hamedi_2017}.
Nonetheless, most of the studies failed to support the long distance propagation of arbitrary shaped optical pulse. 
In this work we investigate the propagation of an arbitrary shaped probe pulse through the three level $\Lambda$ system in presence of a strong continuous control field. Both the probe and control field experience absorption during initial length of propagation. However, the probe pulse is progressively amplified and retains its initial shape. 
The population transfer from ground state to excited state due to control field is the main reason behind the continuous energy transfer to the probe field. With suitable choice of parameters, the probe pulse retains its initial shape and propagates without any delay, distortion and absorption, after the amplification process. 
Conventionally  the gain system manifests an instability due to the interplay between non-linearity and anomalous dispersion \cite{instability1,instability2}, nevertheless the proposed system is immune from instability. The transmitted probe pulse display broadening which is in contrast to the gain associated with pulse narrowing \cite{PhysRevA.64.053809}. \\
The perspective of the current scheme is substantially different from the existing techniques in the following way.
Usually shape preserving, distortion free pulse propagation is associated with a specific temporal shape such as $\sech$, and is generated in a nonlinear medium by balancing the effects of dispersion and self-phase modulation \cite{boyd_chapter_7}. SIT scheme requires the pulse area to be an integral multiple of $2\pi$ to achieve shape preserving, distortion free pulse propagation \cite{McCall1}. Again active Raman gain scheme \cite{arg_1} can be used to amplify a Gaussian pulse as well as produce solitons \cite{arg_3}. Raman scheme requires specific form of the input and also the signal to travel for a longer distance inside the gain media to achieve significant amplification and for the effects of nonlinearity to balance dispersion to produce stable solitons. Finally, chirped pulse amplification (CPA) method can be used for ultra-short pulse amplification by employing clever instrumental techniques that are applied outside the gain medium \cite{Mourou_et_al}. These processes are sensitive to the angle between signal and pump beams for phase-matching geometry. Our proposed scheme is based on intrinsic transient gain phenomena  that can overcome above issues such as fine balance between non-linearity and  dispersion of the medium, specific shape of input envelope, and strict phase-matching constraints,  in order to form amplified stable shape preserving pulses. The broadening of the pulse during the amplification process can be beat by using compressors. Hence our scheme is able to amplify, propagate complicated pulse shapes without any dispersion, absorption and distortion.

The paper is organized as follows. Sec. \ref{sec:theory} contains the theoretical formulation of the relevant level system. Sec. \ref{sec:numerical_results} contains the results of numerical simulations along with detailed explanations. Sec. \ref{sec:summary_and_conclusion} contains the summary and conclusion.

\section{\label{sec:theory}Theoretical formulation}

\subsection{\label{sec:level_system}Level system}
\begin{figure}[h]
\begin{center}
\includegraphics[scale=1]{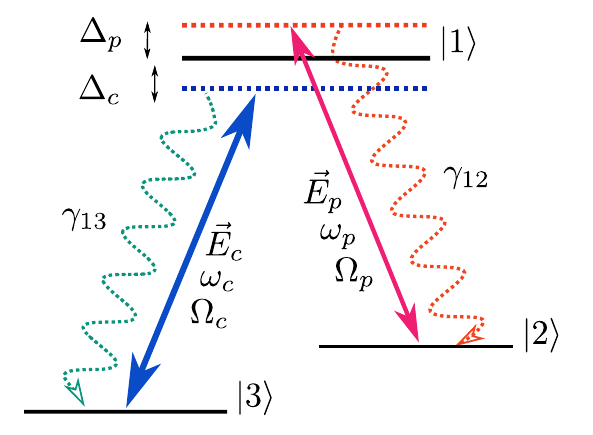}
\vspace*{1mm}
\caption{\label{fig:level_diagram} (Color online) Schematic diagram of a three level $\Lambda$-system. Here, $\ket{1}$ is the excited state, $\ket{2}$ is an intermediate meta stable state and $\ket{3}$ the ground state with energy set to zero. The atomic transitions $\ket{1} \leftrightarrow\ket{2}$ and  $\ket{1}\leftrightarrow\ket{3}$ are coupled by a weak probe field $\vec{E}_p$, with frequency $\omega_p$, and a strong control field $\vec{E}_c$, with frequency $\omega_c$, respectively. The spontaneous emission decay rate of $\ket{1}$ to $\ket{j}$ $(j\in 2,3)$ transition is denoted by $\gamma_{1j}$. The detunings, and Rabi frequencies of the fields are denoted by $\Delta_i$, and $\Omega_i$, respectively ($i \in p,c $ represents probe and control field, respectively).}
\end{center}
\end{figure}

%
A $\Lambda$-system as shown in Fig. \ref{fig:level_diagram} can meet the desired criteria for achieving gain by considering that all population is kept at ground state $\ket{3}$ initially. Unlike the absorption based EIT configuration, in Fig. \ref{fig:level_diagram}, the ground state $\ket{3}$ and excited state $\ket{1}$ are coupled by a strong control field $\vec{E}_c$ and the metastable state $\ket{2}$ and $\ket{1}$ are coupled by a weak probe field $\vec{E}_p$. This configuration leads to population transfer from $\ket{3}$ to $\ket{1}$ which spontaneously decays to $\ket{2}$. The decay from $\ket{1}$ to $\ket{2}$ makes provision for the probe field being resonantly enhanced by stimulated emission in presence of the control field, causing probe amplification. The three level $\Lambda$ system in Fig. \ref{fig:level_diagram} can be realized with the $^{87}$Rb $D_2$ $(5^2S_{1/2} \rightarrow 5^2P_{3/2})$ transition hyperfine structure, by considering the Zeeman sub-levels of the ground state hyperfine levels as, $\ket{3} = \ket{5^2S_{1/2}, F = 1, m_F = -1}$, $\ket{2} = \ket{5^2S_{1/2}, F = 1, m_F = +1}$, and the excited state as $\ket{1}  = \ket{5^2P_{3/2}, F = 0, m_F = 0}$. The probe and control fields are considered to be propagating along $z$ direction and are defined as:

\begin{equation}
\label{eq:fields}
\vec{E}_j (r,t) = \hat{e}_j \mathcal{E}_j(r,t) e^{-i(\omega_j t - \vec{k}_j. \vec{r})} + \text{c.c.},
\end{equation}
where $\hat{e}_j$ are the unit polarization vectors, $\mathcal{E}_j(r,t)$ are the slowly varying envelope functions, $\omega_j$ are the field carrier frequencies and $\vec{k}_j = k_j \hat{z}$ are the wave vectors of the fields. The index $j\in p, c$ represents
probe and control fields, respectively.

The time dependent Hamiltonian of the system, under the electric dipole approximation is given as:

\begin{subequations}
\begin{align}
\bm{H}		&= \bm{H}_0 + \bm{H}_I, \\
\bm{H}_0	&= \hbar (\omega_{13}\ket{1}\bra{1} + \omega_{23}\ket{2}\bra{2}), \\
\bm{H}_I	&= \ket{1}\bra{2} \vec{\bm{d}}_{12}.\hat{e}_p\mathcal{E}_p e^{-i(\omega_p t - k_p z)}, \notag\\
			&+ \ket{1}\bra{3} \vec{\bm{d}}_{13}.\hat{e}_c\mathcal{E}_c e^{-i(\omega_c t - k_c z)} + \mbox{h.c.} ,	
\end{align}
\end{subequations}
where $\omega_{j3}$ $(j \in 1,2)$ denotes the resonance frequency of $\ket{j}\leftrightarrow\ket{3}$ transition and $\vec{\bm{d}}_{j3} = \bra{j}\bm{\hat{d}}\ket{3}$ are matrix elements of the dipole moment operator $\bm{\hat{d}}$, representing the induced dipole moments, corresponding to $\ket{j}\leftrightarrow\ket{3}$ transition. To write the Hamiltonian in a time independent form, the following unitary transformation is used:

\begin{subequations}
\begin{align}
\bm{U} &= e^{-i\bm{V} t},\\
\bm{V} &= \omega_c\ket{1}\bra{1} + (\omega_c - \omega_p)\ket{2}\bra{2}. 
\end{align}
\end{subequations}
The effective Hamiltonian obeying the Schr\"{o}dinger equation in the transformed basis is given as:

\begin{equation}
\mathcal{H} = \bm{U}^{\dagger}.\bm{H}.\bm{U} - \frac{i}{\hbar}\bm{U}^{\dagger}.\frac{\partial \bm{U}}{\partial t},
\end{equation}
which under \textbf{RWA} (Rotating wave approximation) gives

\begin{align}
\label{eq:RWA_hamiltonian}
\frac{\mathcal{H}}{\hbar} &= -\Delta_c\ket{1}\bra{1} - (\Delta_c - \Delta_p)\ket{2}\bra{2} \notag \\
&- \Omega_p\ket{1}\bra{2} - \Omega_c\ket{1}\bra{3} + \text{h.c.}.
\end{align}
The single photon detunings for the probe and control fields are defined as:

\begin{equation}
\Delta_p = \omega_p - \omega_{12} ,\quad \Delta_c = \omega_c - \omega_{13},
\end{equation}
and the Rabi frequencies of probe and control fields are written as:

\begin{equation}
\label{eq:Rabi_frequencies}
\Omega_p = \frac{\vec{\bm{d}}_{12}.\hat{e}_p\mathcal{E}_p }{\hbar}e^{i k_p z}, \quad \Omega_c = \frac{\vec{\bm{d}}_{13}.\hat{e}_c\mathcal{E}_c}{\hbar} e^{ik_c z}.
\end{equation}


The dynamics of atomic state populations and coherences is governed by the following Liouville equation:

\begin{equation}
\frac{\partial \rho}{\partial t} = -i\hbar [\mathcal{H}, \rho] + \mathcal{L}_{\rho}.
\end{equation}
The Liouville operator $\mathcal{L}_{\rho}$, describes all incoherent processes and can be expressed as:

\begin{equation}
\mathcal{L}_{\rho} = -\sum_{i=1}^2 \sum_{\substack{j=1 ,\\ j\neq i}}^3 \frac{\gamma_{ij}}{2}\left(\ket{i}\bra{i}\rho - 2 \ket{j}\bra{j}\rho_{ii} + \rho \ket{i}\bra{i}\right),
\end{equation}
where $\gamma_{ij}$ represent the radiative decay rates from excited
states $\ket{i}$ to ground states $\ket{j}$.

The equations of motion for atomic state populations and coherences of the three level $\Lambda$-system are then given as:

\begin{subequations}
\label{eq:density_matrix_equations}
\small
\begin{align} 
 \dot{\rho}_{11} &= -\Gamma\rho_{11} + i(\Omega_p\rho_{21} - \rho_{12}\Omega^*_p)+ i(\Omega_c\rho_{31} -\Omega^*_c\rho_{13}),\label{eq:density_matrix_equations_1}\\
\dot{\rho}_{12} &= -\bigg(\frac{\Gamma}{2} - i \Delta_p\bigg)\rho_{12} + i\Omega_c\rho_{32} + i\Omega_p(\rho_{22} - \rho_{11}),\label{eq:density_matrix_equations_2}\\
\dot{\rho}_{13} &= -\bigg(\frac{\Gamma}{2} - i\Delta_c\bigg)\rho_{13} + i\Omega_c(\rho_{33}-\rho_{11}) + i\Omega_p\rho_{23},\label{eq:density_matrix_equations_3}\\
\dot{\rho}_{22} &= \gamma_{12}\rho_{11} + i(\Omega^*_p\rho_{12} - \rho_{21}\Omega_p),\label{eq:density_matrix_equations_4}\\
\dot{\rho}_{23} &= -\bigg[\gamma_{23} - i(\Delta_c - \Delta_p)\bigg]\rho_{23} + i(\Omega^*_p\rho_{13} -\Omega_c\rho_{21}),\label{eq:density_matrix_equations_5}\\
\dot{\rho}_{33} &= \gamma_{13}\rho_{11} + i(\rho_{13}\Omega^*_c - \Omega_c\rho_{31}),\label{eq:density_matrix_equations_6}\\
\rho^*_{ij} &= \rho_{ji}\label{eq:density_matrix_equations_7},
\end{align}
\end{subequations}
with initial conditions
\begin{equation}
\rho_{11}(z,0) = \rho_{22}(z,0)= 0\;, {\text {and}}~~\rho_{33}(z,0) = 1.\notag
\end{equation}	
Assuming that the system is closed, the total population remains conserved; {\it i.e.}, $\rho_{11} +\rho_{22} + \rho_{33} = 1$. In Eq. (\ref{eq:density_matrix_equations}), the overdots stand for time derivatives and ``$*$" denotes complex conjugate. The decoherence rate of $\rho_{23}$  is denoted by $\gamma_{23}$ and the total decay rate of excited state $\ket{1}$ is written as $\Gamma = \gamma_{12} +\gamma_{13}$. The decay rate of excited state $\ket{1}$, to states $\ket{2}$ and $\ket{3}$ are assumed to be equal; i.e., $\gamma_{12} = \gamma_{13} = \gamma$.



\subsection{\label{sec:propagation_equations}Propagation equations}
In order to explore the effects of populations and coherences on the propagation dynamics of probe pulse through the gain medium, the study of Maxwells equation is inevitable. Under the slowly varying envelope approximation, the propagation equations for the probe and control fields can be expressed as

\begin{subequations}
\label{eq:Maxwell_equations}
\begin{align}
\bigg(\frac{\partial}{\partial z} + \frac{1}{c}\frac{\partial}{\partial t}\bigg)\Omega_p(z,t) &=\eta_p \rho_{12}(z,t), \label{eq:Maxwell_equations_1}\\
\bigg(\frac{\partial}{\partial z} + \frac{1}{c}\frac{\partial}{\partial t}\bigg)\Omega_c(z,t) &=\eta_c \rho_{13}(z,t),\label{eq:Maxwell_equations_2}
\end{align}
\end{subequations}
where $\eta_i$ $(i\in p,c)$ are called the coupling constants of the respective fields. For simplicity, the resonance frequencies of $\ket{1}\rightarrow\ket{j}$ $(j\in 2,3)$ transitions are assumed to be equal; {\it i.e.}, $\omega_{12}=\omega_{13}$. This gives $\eta_p=\eta_c= \eta$ with $\eta = 3N\lambda^2\gamma/8\pi$, where $N$ is the number of atoms per unit volume inside the medium and $\lambda$ is the wavelength of the fields. To facilitate numerical integration of Eq. (\ref{eq:Maxwell_equations}), a frame moving at the speed of light in vacuum $c$, is used. The necessary coordinate transformations for that are $\tau = t - z/c$, and $\zeta = z$. This allows for the round bracketed terms of Eq. (\ref{eq:Maxwell_equations}) to be replaced by partial derivatives with respect to the single independent variable $\zeta$.
\begin{figure}[b]
\begin{center}
\includegraphics[scale=1]{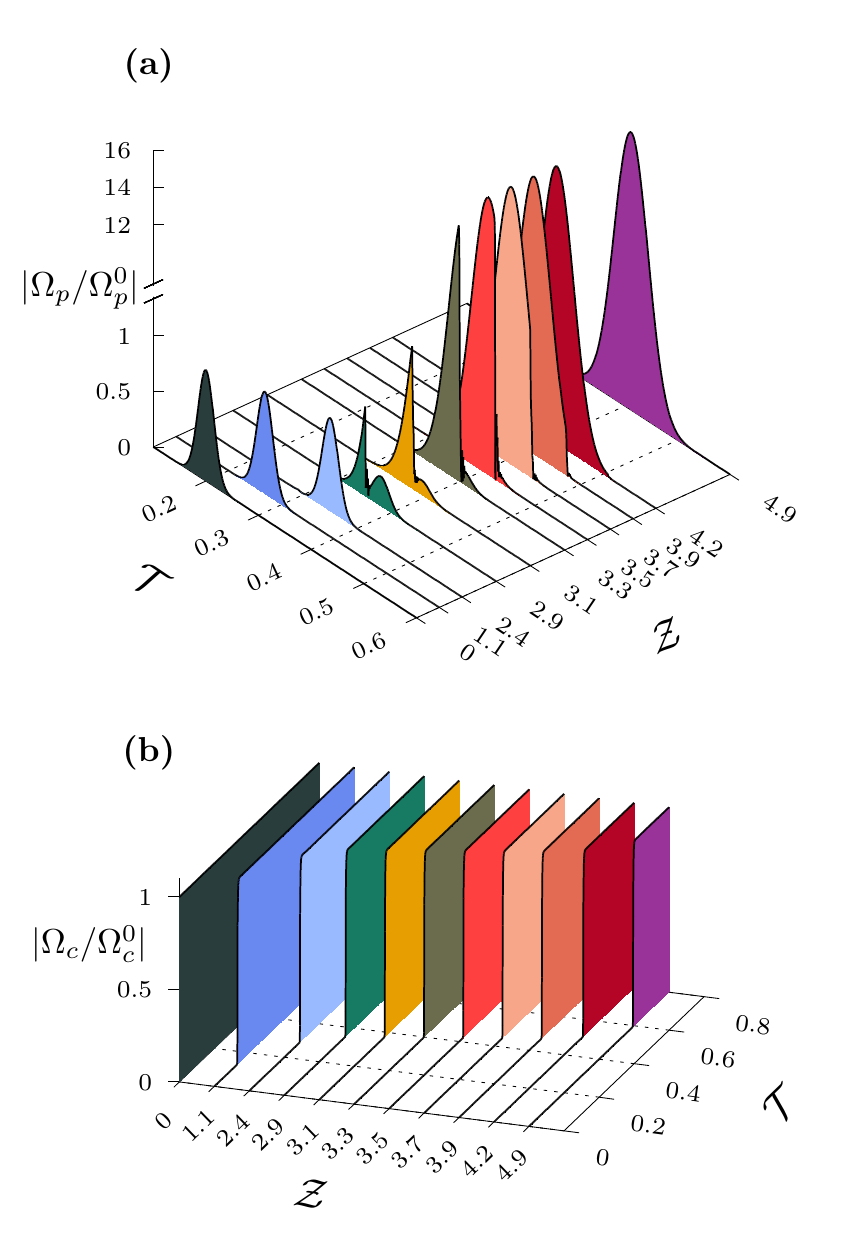}
\vspace*{1mm}
\caption{\label{fig:temporal_profiles}(Color online) (a) Temporal profile of probe pulse at different propagation lengths $\mathcal{Z}$. (b) Temporal profile of control field corresponding to the propagation lengths mentioned in Fig. \ref{fig:temporal_profiles}(a). Parameters used are: $\Omega^0_p = 0.01\gamma$, $\Omega^0_c = 4\gamma$, $\tau_0 = 200/\gamma$, $\sigma_0 = 15/\gamma$ and $\Delta_p = \Delta_c = 0$, $\gamma_{23} = 0.001\gamma$. Field magnitudes are normalized by their respective field magnitudes at $\mathcal{Z} = 0$.  Axes are not made to scale for visibility.}
\end{center}
\end{figure}

\section{\label{sec:numerical_results}Numerical Results}
In this section, the numerical results acquired from coupled Maxwell-Bloch equations are presented. For numerical simulation, we first consider spatiotemporal evolution of a Gaussian probe pulse in the presence of a continuous control field. At the entrance of the medium, the Gaussian probe pulse and the continuous control field  are defined as:
\begin{subequations}
\begin{align}
\Omega_p(0,\tau)	&=	\Omega^0_p \exp \left[-\frac{(\tau-\tau_0)^2}{2\sigma^2_0}\right],\\
\Omega_c(0,\tau)	&=	\Omega^0_c,
\end{align}
\end{subequations}
where, $\Omega^0_p$, $\Omega^0_c$ are the amplitudes of probe and control fields respectively and $\sigma_0$, $\tau_0$ are the initial pulse width and peak position of the probe pulse, respectively. The results are curated in Fig. \ref{fig:temporal_profiles}, where the temporal profiles of probe and control field magnitudes are plotted at different propagation lengths. Time, and propagation length are made dimensionless as $\mathcal{T} = \gamma \tau\footnotesize\times10^{-3}$ and $\mathcal{Z} = \eta\zeta/\gamma\footnotesize\times10^{-3}$ respectively throughout the paper.\\
Both the probe pulse and the continuous control field undergo observable reshaping during propagation. In Fig. \ref{fig:temporal_profiles}(a), from  $0\le\mathcal{Z}\le 2.4$, the probe pulse experiences group delay, absorption and indiscernible broadening with increasing propagation length. From $2.4\le\mathcal{Z} \le4.2$, the probe undergoes substantial reshaping and amplification. Beyond $\mathcal{Z} \ge 4.2$, a broadened and amplified Gaussian probe pulse is obtained, which travels at the speed of light in vacuum, without any delay, absorption and dispersion.\\
In Fig. \ref{fig:temporal_profiles}(b), the control field undergoes absorption at its leading end as it propagates though the medium, giving it an appearance of experiencing delay at the leading end.
A detailed explanation of the results in Fig. \ref{fig:temporal_profiles} are provided in the following sections.

\subsubsection{Absorption of control}
\begin{figure}[H]
\begin{center}
\includegraphics[scale=1]{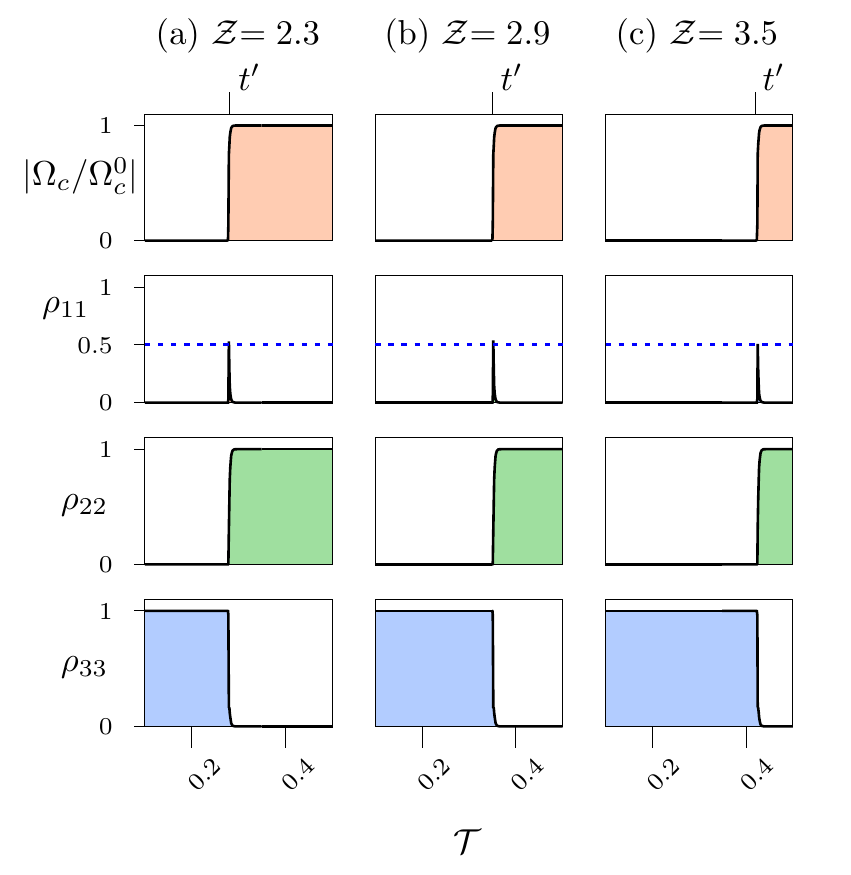}
\caption{\label{fig:2d_population_transfer}(Color online) Temporal profile of control field and population distribution are plotted column wise at propagation lengths $\mathcal{Z} =2.3$, $2.9$, and $3.5$ in (a), (b), (c), respectively. In the topmost row, $t^\prime$ represents the position of control field's leading end on time axis. Control field magnitude normalization and parameters used are same as Fig. \ref{fig:temporal_profiles}. Figures pertaining to a particular row and column have a common vertical and horizontal axes, respectively.}
\end{center}
\end{figure}

\begin{figure}[H]
\begin{center}
\includegraphics[scale=1]{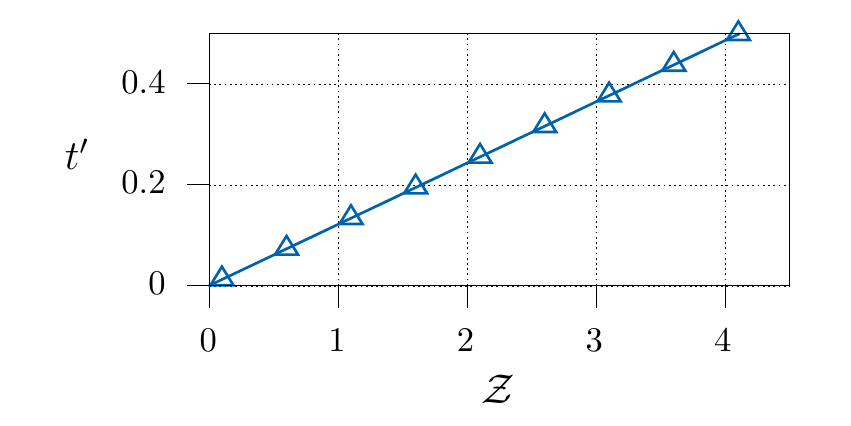}
\caption{\label{fig:control_leading_edge_position_vs_z}Position of control field's leading end $t^\prime$, on time axis vs propagation length $\mathcal{Z}$. The slope of the graph $\beta\approx0.12168$, is the rate at which the control field's leading end appears to move forward along time axis with increasing $\mathcal{Z}$. Parameters used are same as Fig.  \ref{fig:temporal_profiles}.}
\end{center}
\end{figure}

In Fig. \ref{fig:2d_population_transfer}, at the position of control field's leading end on time axis, there occurs a population transfer from $\ket{3}$ to $\ket{2}$, with an intermediary transient population transfer to $\ket{1}$. This population jump from $\ket{3}$ to $\ket{1}$ as shown in Fig. \ref{fig:2d_population_transfer} ($2$\textsuperscript{nd} row), requires a certain amount of control field energy. Thus, for every infinitesimal increment in $\mathcal{Z}$, energy corresponding to a tiny portion of the control field's total area gets absorbed in the medium. Therefore, at any given $\mathcal{Z}$, a chunk of control field area shows up missing from the leading end as shown in Fig. \ref{fig:2d_population_transfer} ($1$\textsuperscript{st} row).
\noindent The control field undergoes linear absorption with increasing propagation length $\mathcal{Z}$, as shown in Fig. \ref{fig:control_leading_edge_position_vs_z}, where the graph between the position of control field's leading end ($t^\prime$) on time axis vs $\mathcal{Z}$ gives a straight line.
The formula for the slope of the graph in Fig. \ref{fig:control_leading_edge_position_vs_z} is derived as follows. The energy density of control field is given as $|\mathcal{E}_c|^2/8\pi$. Therefore, if $A$ is the area of cross section of the control laser beam, then the energy passing through the transverse plane at any given $z$, for a time interval $t^\prime/\gamma$ is
\begin{equation}
\label{eq:energy_transfered}
\int_{0}^{t^\prime/\gamma} \frac{cA|\mathcal{E}_c|^2}{8\pi}d\tau,
\end{equation}
which is also the amount of control field energy that gets absorbed after propagating for a distance $z$, inside the medium. Again, due to control field absorption, approximately half of the population momentarily transfer from ground state $\ket{3}$ to excited state $\ket{1}$ as marked in Fig. \ref{fig:2d_population_transfer} ($2$\textsuperscript{nd} row). Therefore, when each atom in a homogeneous  medium with atomic number density $N$, absorbs one control field photon of energy $\hbar \omega_{13}$, then the total energy absorbed by the medium is
\begin{equation}
\label{eq:energy_absorbed}
E \approx \hbar\omega_{13} \frac{N}{2} A z.
\end{equation}
Equating Eqs. (\ref{eq:energy_transfered}) and (\ref{eq:energy_absorbed}) gives
\begin{equation}
t^\prime \approx  \frac{2}{|\Omega_c/\gamma|^2}\mathcal{Z} \label{eq:theoretical_slope}.
\end{equation}
\noindent Therefore, the theoretical slope of $t^\prime$ vs $\mathcal{Z}$ plot is $\beta_t = 2/|\Omega_c/\gamma|^2$. The slopes obtained numerically ($\beta_n$) for different control magnitudes are in good agreement with the theoretical ones as shown in Table \ref{tab:beta}. Where, the ratio $\beta_t/\beta_n\approx 1$. 
\begin{table}[H]
\caption{\label{tab:beta} Table showing comparison between $\beta_n$ and $\beta_t$}
\begin{ruledtabular}
\begin{tabular}{ccddd}
\multicolumn{1}{c}{\textrm{$|\Omega_c/\gamma|$}}&
\multicolumn{1}{c}{\textrm{$\beta_n$}}&
\multicolumn{1}{c}{\textrm{$\beta_t$}}&
\multicolumn{1}{c}{\textrm{$\beta_t/\beta_n$}}\\
\hline
2 & 0.492 & 0.5 	& 1.02	\\  
3 & 0.219 & 0.22 	& 1.00	\\
4 &	0.123 &	0.125 	& 1.02	\\
5 &	0.077 &	0.08 	& 1.04	\\
6 &	0.053 &	0.056  	& 1.06	\\
7 &	0.039 &	0.041  	& 1.05
\end{tabular}
\end{ruledtabular}
\end{table}
\subsubsection{\label{sec:probe_dispersion}Probe delay, dispersion and absorption}
\begin{figure}[H]
\begin{center}
\includegraphics[scale=1]{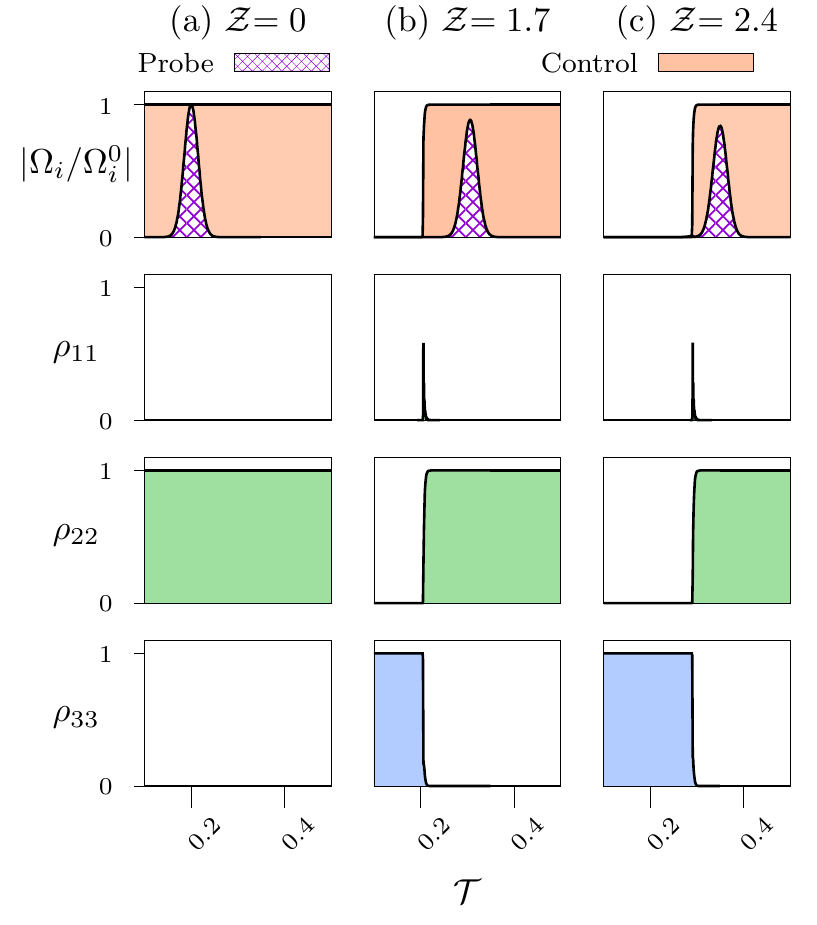}
\caption{\label{fig:2d_population_distribution_before_overtake}(Color online) ($1$\textsuperscript{st} row) Probe ($i = p$, crisscross) and control ($i = c$, solid) field magnitude vs time $\mathcal{T}$, are plotted column wise for $\mathcal{Z}=0$, $1.7$, and $2.4$ in (a), (b), (c) respectively. (Other rows) Population distributions at the corresponding propagation lengths. The probe is shown traveling inside a medium with population distribution $\rho_{11} = \rho_{33} = 0$, and  $\rho_{22} = 1$. Normalization of field magnitudes and parameters used are same as Fig. \ref{fig:temporal_profiles}. Figures connected to a particular row and column have a common vertical and horizontal axes, respectively.}
\end{center}
\end{figure}

The probe pulse travels incontrovertibly under the influence of control field from $0\le\mathcal{Z}\le2.4$ as shown in Fig. \ref{fig:2d_population_distribution_before_overtake} ($1$\textsuperscript{st} row). The presence of control field creates a population distribution of $\rho_{11} = \rho_{33} = 0, \rho_{22} = 1$ as shown in Fig. \ref{fig:2d_population_distribution_before_overtake} ($2$\textsuperscript{nd}, $3$\textsuperscript{rd}, $4$\textsuperscript{th} row), which serves as an absorbing, dispersive medium for the probe pulse. Hence, the probe undergoes noticeable absorption and delay within $0\le\mathcal{Z}\le2.4$. The probe absorption coefficient $\alpha^\prime$, can be calculated from the imaginary part of the probe susceptibility $\chi$, of the medium. The probe susceptibility $\chi$, is written in terms of the atomic coherence as:
\begin{subequations}
\begin{align}
&\chi = \frac{\eta^\prime}{\Omega_p}\rho_{12} \quad \left(\eta^\prime = \frac{3N\lambda^3\gamma}{32\pi^3}\right),\label{eq:susceptibility} \\
&\rho_{12} = \frac{-\Omega _p \left[|\Omega _p|^2+\gamma (\gamma _{23}-i \Delta _p)\right]}{(\gamma -i\Delta _p)\left[\gamma  (\Delta _p+i \gamma_{23})+i |\Omega _p|^2\right]+i\gamma  | \Omega _c| ^2},\label{eq:rho_12}		
\end{align}
\end{subequations}	
where $\rho_{12}$, in Eq. (\ref{eq:rho_12}), comes from the steady state solutions of Eq. (\ref{eq:density_matrix_equations}) with $\Delta_c = 0$, under the weak probe $(\Omega^0_p \ll \Omega^0_c)$ approximation. The steady state boundary condition $\rho_{11} = \rho_{33} = 0$, and $\rho_{22} = 1$ is used for the derivation of Eq. (\ref{eq:rho_12}). The probe absorption coefficient is defined as $\alpha^\prime = 2\pi\omega\text{Im}[\chi]/c$, which is made dimensionless as $\alpha^{(d)} = \gamma\alpha^\prime/\eta$. At two photon resonance $(\Delta_p = \Delta_c = 0)$,
\begin{equation}
\label{eq:probe_absorption_coefficient}
\alpha^{(d)} = 1-\frac{|\Omega_c|^2}{\gamma \gamma _{23}+|\Omega_c|^2+|\Omega_p|^2}.
\end{equation}
For the parameters in Fig. \ref{fig:temporal_profiles}, $\alpha^{(d)} \approx 6.9\times10^{-5}$, which is close to the value obtained numerically, $6.5\times10^{-5}$.
It is well known that, after propagating for a distance $z$ inside an absorbing, dispersive medium, a Gaussian pulse of the form:
\begin{equation}
\Omega_p(0,t) = \Omega^0_p \exp \left[\frac{-t^2}{2\sigma_0^2}\right],\label{eq:E(0,t)}
\end{equation}
gets modified as:
\begin{equation}
\label{eq:E(z,t)}
\Omega_p(z,t) = \Omega^0_p\sqrt{\frac{\sigma_0^2}{\sigma_0^2 - ik^{''}_0z}}\exp\left[\frac{-(t - k^{\prime}_0 z)^2}{2(\sigma_0^2 - ik^{''}_0 z)}-\alpha^\prime z\right],
\end{equation}
where $k^{\prime}_0 = \frac{\partial k(\omega)}{\partial \omega}\big|_{\omega_0}$ and $k^{''}_0 =\frac{\partial^2 k(\omega)}{\partial \omega^2}\big|_{\omega_0}$ with $k(\omega)$ and $\omega_0$ being the propagation vector and carrier frequency of the field respectively. Equation (\ref{eq:E(z,t)}) shows that the Gaussian pulse gets delayed by a time $\tau_d = k^{\prime}_0 z$ and the pulse width gets modified as $\sigma^2(z) = \sigma_0^2 - ik^{''}_0 z$, after traveling a distance $z$ inside any dispersive medium. The parameters $k^{\prime}_0$ and $k^{\prime\prime}_0$ are written in terms of the probe susceptibility $\chi$, as:
\begin{equation}
\label{eq:k_prime_0_&_k_dprime_0}	
k^{\prime}_0 = \left(1+2\pi\omega \frac{\partial \chi}{\partial \omega}\bigg|_{\omega_0}\right)/c,\quad
k^{''}_0 = \frac{2\pi \omega_0}{c}\frac{\partial^2\chi}{\partial\omega^2}\bigg|_{\omega_0},
\end{equation}
which are made dimensionless as $\kappa_1 = \gamma^2 k^\prime_0/\eta$ and $\kappa_2 = \gamma^3 k^{\prime\prime}_0/\eta$ respectively. Substituting Eqs. [(\ref{eq:susceptibility}), (\ref{eq:rho_12})] in Eq. (\ref{eq:k_prime_0_&_k_dprime_0}) gives:
\begin{subequations}
\begin{align}
\kappa_1 =&  \frac{\gamma^2 \abs{\Omega _c}^2-(\gamma \gamma _{23}+ \abs{\Omega _p}^2)^2}{(\gamma  \gamma _{23}+ \abs{\Omega_c}^2 + \abs{\Omega _p}^2)^2},\label{eq:k_prime_0_expression}\\
\kappa_2 =& 2 i \{\gamma ^2 \abs{\Omega _c}^2[\gamma  (2 \gamma _{23}+\gamma )+2 \abs{\Omega _p}^2]\notag \\
&-(\gamma  \gamma_{23}+  \abs{\Omega _p}^2)^3\}/(\gamma  \gamma _{23}+ \abs{\Omega _c}^2+ \abs{\Omega _p}^2)^3.\label{eq:k_dprime_0_expression}
\end{align}
\end{subequations}
For the parameters in Fig. \ref{fig:temporal_profiles}, Eq. (\ref{eq:k_prime_0_expression}) gives $\kappa_1= 0.0625$, which is close to the numerical value, $0.0623$. The theoretical value of $\kappa_2= 7.83\times 10^{-3} i$, for parameters in Fig. \ref{fig:temporal_profiles}, is very small to cause any significant change in the pulse width $\sigma(z) = \sqrt{\sigma_0^2 - ik^{''}_0 z}$. This is why in Fig. \ref{fig:2d_population_distribution_before_overtake} ($1$\textsuperscript{st} row), from $0\le\mathcal{Z} \le2.4$ no noticeable change in the pulse width is observed.

The parameter $\kappa_1$ can be perceived as the rate at which the probe pulse proceeds along time axis with increasing $\mathcal{Z}$. The aforementioned value of $\kappa_1= 0.0625$, obtained using Eq. (\ref{eq:k_prime_0_expression}), is less than $\beta = 0.125$ (see Table \ref{tab:beta}). Therefore, the control field's leading end moves forward along time axis at a greater rate than the probe pulse as evident from Fig. \ref{fig:2d_population_distribution_before_overtake} ($1$\textsuperscript{st} row). 

In Fig. \ref{fig:2d_population_distribution_before_overtake} ($3$\textsuperscript{rd} column), beyond $\mathcal{Z}\geq 2.4$, control field's leading end begins to overlap with the probe pulse, thereby initiating the process of pulse reshaping and amplification, which continues till the control field's leading end completely overtakes the probe pulse at $ \mathcal{Z}= 4.2$. The process of probe amplification is explained in the next section. 
%


\subsubsection{\label{sec:probe_amplification}Probe reshaping and amplification}

\begin{figure}[H]
\begin{center}
\includegraphics[scale=1]{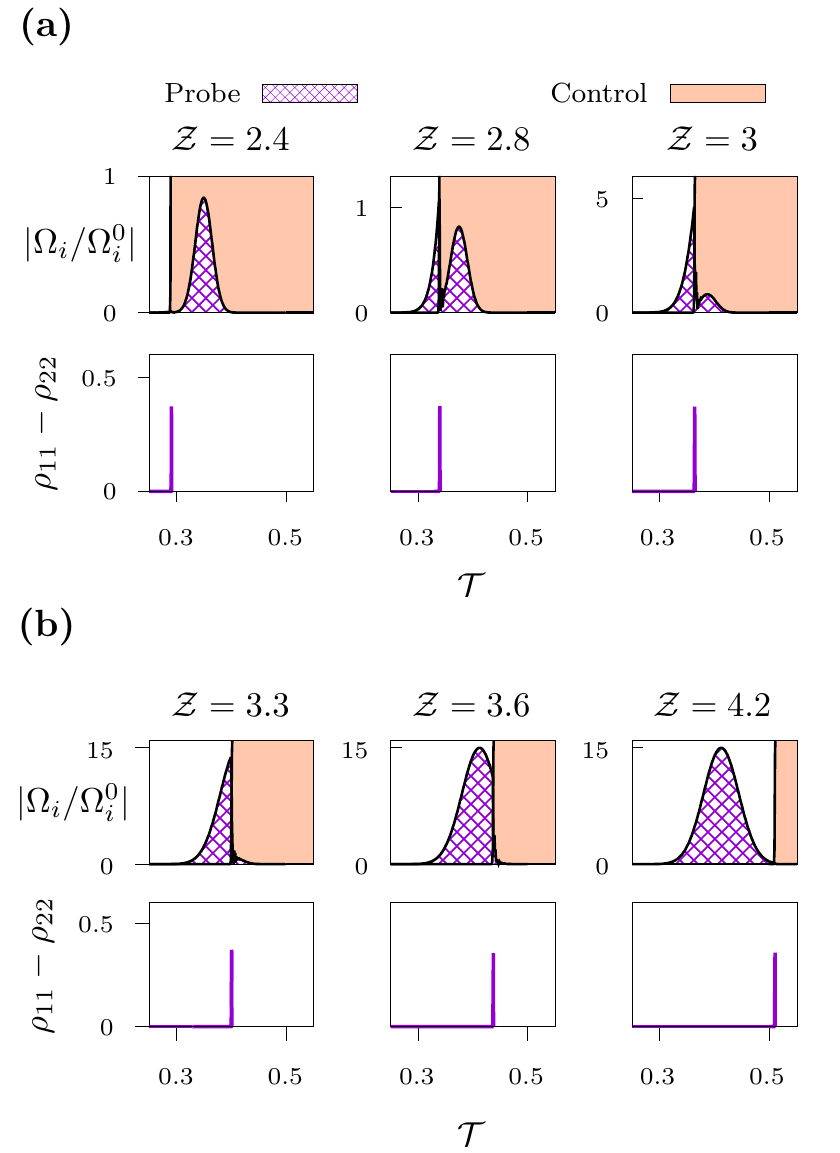}
\caption{\label{fig:probe_amplification}(Color online) [$1$\textsuperscript{st} row of (a), (b)] Temporal profiles of probe ($i = p$, crisscross) and control ($i = c$, solid) field magnitude are plotted column wise at different propagation lengths $\mathcal{Z}$. [$2$\textsuperscript{nd} row of (a), (b)] ($\rho_{11} - \rho_{22}$) vs time $\mathcal{T}$, at the corresponding propagation lengths. Plot illustrates how the positive peak of ($\rho_{11} - \rho_{22}$) at the position of control field's leading end causes probe amplification at every $(\zeta,\tau)$. Normalization of field magnitude and the parameters used are same as Fig. \ref{fig:temporal_profiles}, except here the normalized control field magnitude is upscaled in a way to accommodate the probe pulse inside it, indicating $\Omega^0_p \ll \Omega^0_c$. In both Figs. \ref{fig:probe_amplification}(a), (b), plots connected to a particular column have a common time axis. The vertical axis for $\left(\rho_{11} - \rho_{22}\right)$ plots remain same along a row while the plots for field magnitudes have different vertical axis along a row.}
\end{center}
\end{figure}

As seen earlier in Fig. \ref{fig:2d_population_transfer} ($2$\textsuperscript{nd}  row), there occurs a momentary population transfer from the ground state $\ket{3}$ to excited state $\ket{1}$ at the position of control field's leading end. This leads to a positive peak in $\left(\rho_{11} - \rho_{22}\right)$ causing population inversion in the $\ket{1}\leftrightarrow\ket{2}$ channel [see Figs. \ref{fig:probe_amplification}(a), (b)]. Hence, in presence of the probe field there is stimulated emission of probe photons in the $\ket{1}\leftrightarrow\ket{2}$ channel causing probe amplification. This is illustrated in Fig. \ref{fig:probe_amplification}, where the temporal profiles of probe and control field magnitudes along with $\left(\rho_{11}- \rho_{22}\right)$ are plotted column wise at different propagation lengths. 

In Fig. \ref{fig:probe_amplification}, with increasing $\mathcal{Z}$, both the positive peak of $(\rho_{11} - \rho_{22})$ and control field's leading end, proceed collinearly at a greater rate than the probe pulse along time axis. At $\mathcal{Z}= 2.4$, the leading end of both probe and control fields coincide on the time axis. From $2.4\le\mathcal{Z} \le 4.2$, the control field's leading end or in other words, the positive peak of $(\rho_{11} - \rho_{22})$ grazes through the probe pulse on time axis. This causes probe amplification at the position of the positive peak of $(\rho_{11} - \rho_{22})$ on time axis for every space time coordinate.

In Fig. \ref{fig:probe_amplification}(b) ($3$\textsuperscript{rd} column), beyond $\mathcal{Z}\geq 4.2$, both the positive peak of $(\rho_{11} - \rho_{22})$ and the control field's leading end completely overtakes the probe pulse, halting further amplification. The probe pulse then travels at the speed of light in vacuum, without any delay, distortion and absorption, whilst retaining its initial Gaussian shape but with a larger pulse width. This stable propagation of the probe pulse is explained in the next section.
 
%
%
%
\subsubsection{\label{sec:stable_pulse_propagation}Stable probe propagation}

\begin{figure}[H]
\begin{center}
\includegraphics[scale=1]{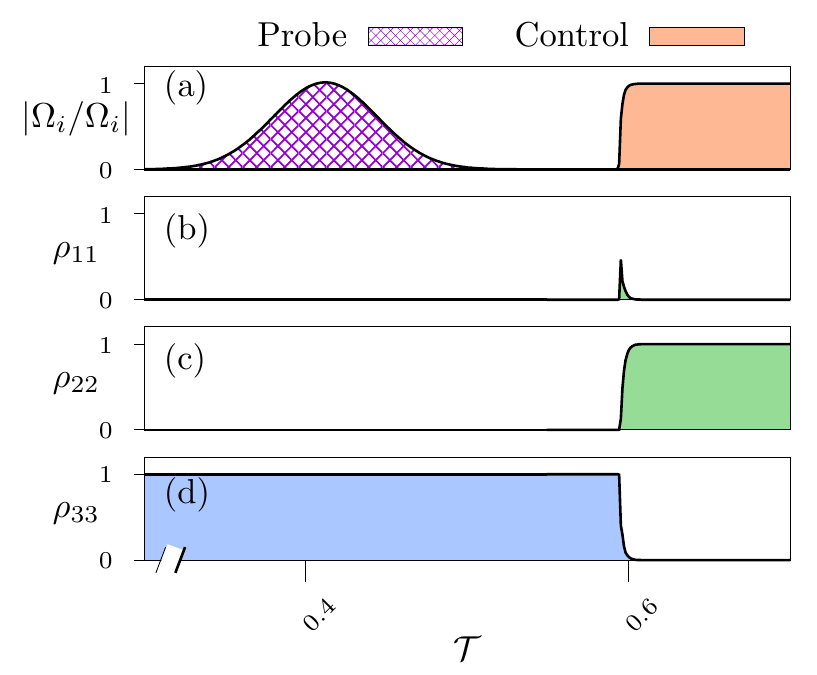}
\caption{\label{fig:analysis_after_crossover}(Color online) (a) Temporal profiles of probe (crisscross) and control (solid) field magnitudes at $\mathcal{Z}= 4.9$. (b), (c), (d) Population distribution of $\ket{1}$, $\ket{2}$ and $\ket{3}$ respectively. The plot illustrates the population distribution $\rho_{11} = \rho_{22} = 0, \rho_{33} = 1$ as seen by the probe after the control field's leading end completely overtakes it. Both probe and control field magnitudes are normalized to unity. Parameters used are same as Fig. \ref{fig:temporal_profiles}. All figures have a common time axis.}
\end{center}
\end{figure}

The technique of pulse amplification at hand, ensures a stable pulse propagation upon completion of amplification. This is made possible due to the population distributions created by the control field as it propagates through the medium. Figure \ref{fig:analysis_after_crossover} shows the temporal profiles of the fields along with population distribution at a position $\mathcal{Z}= 4.9$ after the completion of amplification process. In Fig. \ref{fig:analysis_after_crossover}, within $0\leq\mathcal{T}\leq 0.6$, the probe pulse sees a population distribution $\rho_{11} = \rho_{22} = 0, \rho_{33} = 1$ [see Figs. \ref{fig:analysis_after_crossover}(b), (c), (d)], {\it i.e.}, all population gets settled in the dark state  $\ket{3}$ \cite{eit_1}. Therefore, the general state $\ket{\psi}$ of the system within the time interval $0\le\mathcal{T}\le0.6$ can be written as:
\begin{equation}
\label{eq:general_state}
\ket{\psi} = a_1\ket{1} +a_2\ket{2}+a_3\ket{3} = \ket{3},
\end{equation}
where $a_i$ $(i = 1,2,3)$ represent the probability amplitudes of state $\ket{i}$. Equation (\ref{eq:general_state}) implies $a_1 = a_2  = 0, a_3  = 1$. Therefore, from the definition of density matrix elements $\rho_{12} = a_1^*a_2 = 0$. The propagation equation for the probe pulse [see Eq. (\ref{eq:Maxwell_equations_1})] then becomes:

\begin{align}
\frac{\partial \Omega_p(\zeta,\tau)}{\partial \zeta} &= i\eta\rho_{12}(\zeta,\tau)\notag \\
&= 0,
\end{align}
which resembles a free space propagation equation. Thus at the end of amplification process, the probe pulse travels freely, unattenuated and undistorted at the speed of light in vacuum, even in the presence of medium.
%
%
%
\subsubsection{\label{sec:pulse_broadening}Probe pulse broadening}
	
\begin{figure}[H]
\begin{center}
\includegraphics[scale=1]{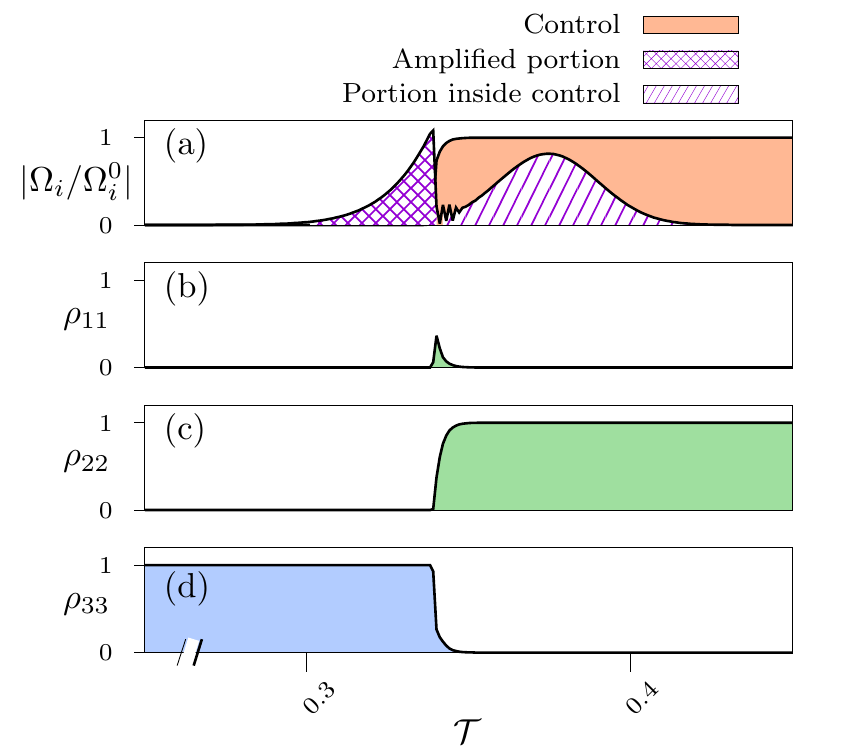}
\caption{\label{fig:pulse_broadening_explaination}(Color online) (a) Temporal profiles of probe and control field magnitudes at $\mathcal{Z} = 2.8$. (b), (c), (d) Population distribution of $\ket{1}$, $\ket{2}$ and $\ket{3}$ respectively. The plot shows how the portion of the probe pulse behind the control field's leading end [crisscross pattern in (a)] sees a population distribution $\rho_{11} = \rho_{22} = 0$, $\rho_{33} = 1$. Whereas, the portion of probe pulse inside the control field [oblique line pattern in (a)] sees a different population distribution $\rho_{11} = \rho_{33} = 0$, $\rho_{22} = 1$. The former behaves like dark state, while the later serves as a bright state. Hence the oblique line portion continues to experience delay, absorption and dispersion, while the crisscross portion undergoes free space propagation. Normalization of field magnitudes and parameters used are same as Fig. \ref{fig:temporal_profiles}. All figures have a common time axis.}
\end{center}
\end{figure}

Apart from the negligible pulse broadening due to dispersion in the region $0\le\mathcal{Z}\le2.4$ [Fig. \ref{fig:2d_population_distribution_before_overtake} ($1$\textsuperscript{st} row)], the probe pulse is subjected to a noticeable pulse broadening during amplification within $2.4\le\mathcal{Z}\le3.9$ [Fig. \ref{fig:probe_amplification}(b) ($1$\textsuperscript{st} row)]. This section provides a qualitative explanation for such a pulse broadening. Figure \ref{fig:pulse_broadening_explaination}, shows the temporal profiles of the fields along with population distribution at a position amidst the amplification process. During the amplification process, the probe pulse sort of splits into two portions as indicated by the crisscross and oblique line patterns in Fig. \ref{fig:pulse_broadening_explaination}(a). The crisscross pattern indicates the amplified portion of probe and the oblique line pattern represents the portion of probe which is yet to be amplified. In Figs. \ref{fig:pulse_broadening_explaination}(b), (c), (d), the oblique line portion sees a population distribution $\rho_{11}=\rho_{33} = 0, \rho_{22} = 1$, and the crisscross portion sees a different population distribution $\rho_{11} = \rho_{22} = 0,\rho_{33} = 1$. The former serves as an absorbing, dispersive medium, while the later behaves like dark state, as discussed earlier in Sec. \ref{sec:probe_dispersion} and \ref{sec:stable_pulse_propagation}, respectively. Therefore, the crisscross portion propagates freely without experiencing any delay, distortion and absorption. But the oblique line portion continues to experience noticeable delay, with small absorption and insignificant broadening due to dispersion. It's due to this delay, experienced by the oblique line portion, the probe pulse gets elongated as the control field grazes through it along time axis, with increasing propagation length. The small absorption experienced by the oblique line portion is compensated by amplification and doesn't lead to any noticeable shape distortion.   

%

\subsubsection{\label{sec:pulse_width_calculation}Evaluation of probe pulse width}

\begin{figure}[h]
\centering
\includegraphics[scale=1]{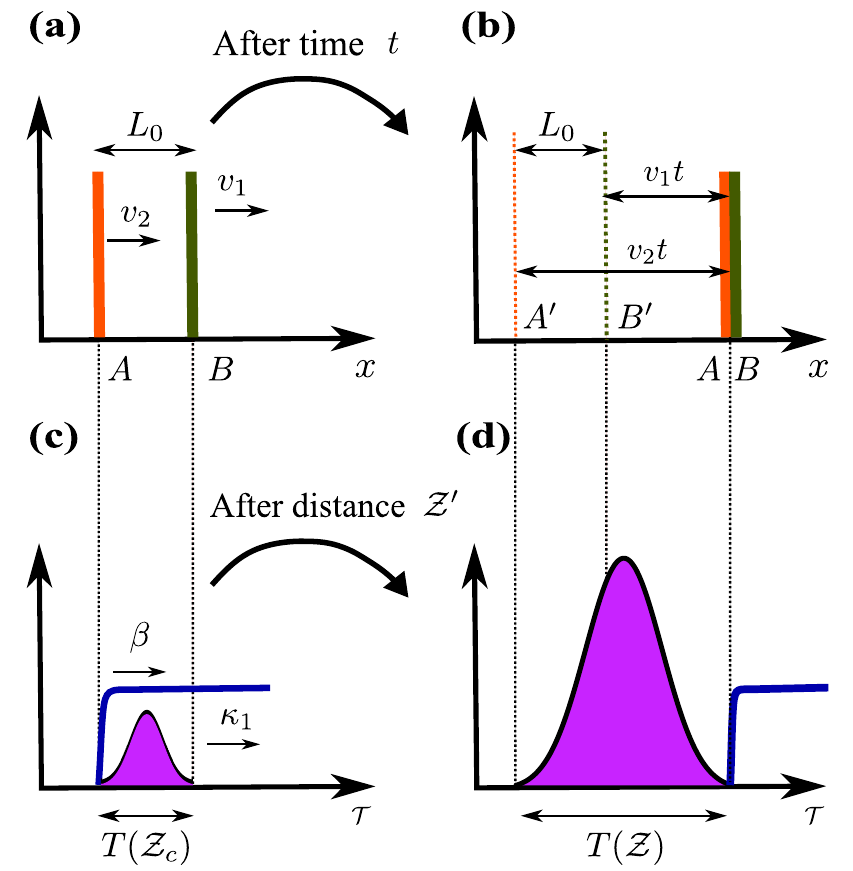}
\vspace*{-1.5mm}
\caption{\label{fig:analogy}(Color online) A diagram illustrating an analogy of the problem to a similar classical problem.}
\end{figure}

To check the predictability of the system, a formula for the probe pulse width at the end of amplification process is derived in terms of experimentally available parameters. To simplify the calculation an analogy to a simple classical problem is drawn in Fig. \ref{fig:analogy}. 

In Fig. \ref{fig:analogy}(a), the bars indicated by ``A" and ``B" move forward along $x$ axis at rates $v_1$ and $v_2$ respectively ($v_2>v_1$). The bars ``A" and ``B" coincide in an elapsed time $t$ after traveling a distance $A^{\prime}A$ and $B^{\prime}B$ respectively [Fig. \ref{fig:analogy}(b)]. In Fig. \ref{fig:analogy}(c), the trailing end of probe pulse and the control field's leading end move forward along time axis at rates $\beta$ and $\kappa_1$ per unit propagation length $\mathcal{Z}$, respectively ($\beta >\kappa_1$). By analogy the control field's leading end is represented by the bar ``A" and, the trailing end of probe pulse is represented by the bar ``B". In Fig. \ref{fig:analogy}(c), $\mathcal{Z}_c$ denotes the propagation length at which both the leading end of probe and control coincide on time axis.
In Fig. \ref{fig:analogy}(d), after propagating a distance $\mathcal{Z^\prime}$, the control field's leading end completely overtakes the probe, leaving behind a broadened, amplified Gaussian probe pulse. From observation, the full extent of the probe pulse $T(\mathcal{Z})$ [Fig. \ref{fig:analogy}(d)], after the amplification process, is simply the time taken by control field's leading end to overtake the trailing end of probe pulse on time axis. 

From analogy, $T(\mathcal{Z})$ is equivalent to $A^\prime A = v_2t$ in Fig. \ref{fig:analogy}(b). From Fig. \ref{fig:analogy}(b), with simple manipulation $A^\prime A$ can be written as:

\begin{align}
\label{eq:A_Aprime}
A^\prime A = \frac{v_2 L_0}{(v_2 - v_1)}.
\end{align}
From analogy, in Fig. \ref{fig:analogy},
\begin{equation*}
A^\prime A  \equiv T(\mathcal{Z}), \quad v_2 \equiv  \beta, \quad v_1 \equiv \kappa_1,\quad L_0 \equiv T(\mathcal{Z}_c).
\end{equation*}%
Therefore, Eq. (\ref{eq:A_Aprime}) gives
\begin{equation}
\label{eq:T(Z)_expression}
T(\mathcal{Z}) = \frac{\beta T(\mathcal{Z}_c)}{(\beta - \kappa_1)},
\end{equation}
where, with some simple algebra, $\mathcal{Z}_c$ can be written as:

\begin{align}
\label{eq:Z_c}
\mathcal{Z}_c &= \frac{-B - \sqrt{B^2 - 4C}}{2},\\
B &= -\frac{8(\beta - \kappa_1)\gamma\tau_0 + \kappa_2 a}{4(\beta - \kappa_1)^2}\mbox{\quad $\left[a = 2\sqrt{2\ln(10^3)}\right]$}, \notag\\
C &= \frac{4\gamma^2\tau_0^2 - \gamma^2\sigma_0^2 a^2}{4(\beta - \kappa_1)^2}\notag.
\end{align}
The pulse width $\sigma(\mathcal{Z})$, can be calculated by using the relation (see appendix \ref{app:A}):
\begin{equation}
\label{eq:relation_between_T_&_sigma}
T(\mathcal{Z}) = 2\sqrt{2\ln(10^3)} \sigma(\mathcal{Z}),
\end{equation}
where $\sigma(\mathcal{Z})^2 = \sigma_0^2 - i\kappa_2\mathcal{Z}$. 
Substituting the earlier obtained values of $\kappa_1 = 0.06233$, $\kappa_2 = 7.83\times10^{-3}$ (see Sec. \ref{sec:probe_dispersion}) and $\beta= 0.123$ (see Table \ref{tab:beta}) in Eqs. [(\ref{eq:T(Z)_expression}) - (\ref{eq:relation_between_T_&_sigma})] gives $T(\mathcal{Z}) = 235.099$, which is close to the numerically obtained value, $237.545$. 

As seen in Eqs. [(\ref{eq:theoretical_slope}), (\ref{eq:k_prime_0_expression}), (\ref{eq:k_dprime_0_expression})] the value of $\beta$ and $\kappa_1$, $\kappa_2$ (for $\Omega^0_p \ll \Omega^0_c$) are merely dependent on $|\Omega_c|$ and independent of the initial pulse width $\sigma_0$, which makes $T(\mathcal{Z})$ [Eq. (\ref{eq:T(Z)_expression})] independent of $\sigma_0$. Therefore, for a particular control field intensity, the ratio $T(\mathcal{Z})/T(0)$ is expected to remain constant for any arbitrary $\sigma_0$.
\begin{table}[ht]
\caption{\label{tab:ratio} Table shows the ratio of $T(\mathcal{Z})$, to the initial full extent of probe pulse $T(0)$.}
\begin{ruledtabular}
\begin{tabular}{ccddd}
\multicolumn{1}{c}{\textrm{$\sigma_0\gamma$}}&
\multicolumn{1}{c}{\textrm{$\abs{\Omega^0_p/\gamma}$}}\footnote{The values of this column are to be multiplied by $10^{-3}$.}&
\multicolumn{1}{c}{\textrm{$T(8.9)$}}&
\multicolumn{1}{c}{\textrm{$T(0)$}}&
\multicolumn{1}{c}{\textrm{$T(8.9)/T(0)$}}\\
\hline
15    & 10  & 237.545 & 111.51 & 2.13 \\
30    & 7.071 & 463.522 & 223.02 & 2.078\\
45    & 5.774 & 693.604 & 334.53 & 2.073\\
60    & 5 & 919.856 & 446.04 & 2.062\\
\end{tabular}
\end{ruledtabular}
\end{table}
Keeping $\Omega^0_c = 4\gamma$ and area of probe pulse constant, the ratios $T(\mathcal{Z})/T(0)$, obtained numerically for different values of $\sigma_0$ are tabulated in Table \ref{tab:ratio}. In Table \ref{tab:ratio}, the ratio $T(\mathcal{Z})/T(0) \approx 2$, remains constant for different initial probe pulse width $\sigma_0$, which matches the expected result. Therefore, the formula in Eq. (\ref{eq:T(Z)_expression}) correctly estimates the probe pulse width at the end of amplification process, countenancing the correctness of the numerical simulation.

\subsubsection{\label{sec:ancillary_results}Ancillary results}

\begin{figure}[H]
\begin{center}
\includegraphics[scale=1]{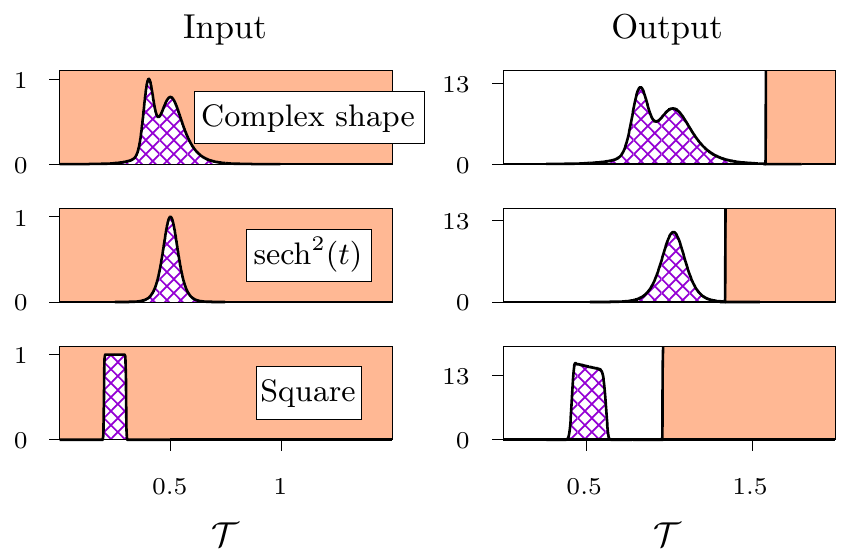}
\vspace*{0.1pt}
\caption{\label{fig:ancillary_results}(Color online) (left column) Temporal profiles of probe (crisscross) and control (solid) field magnitude at $\mathcal{Z} = 0$, for different probe pulse shapes. (right column) Corresponding temporal profiles of probe and control field magnitudes at the end of amplification process. Field magnitude scaling, normalization are same as Fig. \ref{fig:probe_amplification} and parameters used are same as Fig. \ref{fig:temporal_profiles}, with the additional parameters: \footnotesize$\sigma_1 = 30/\gamma$, \footnotesize$\sigma_2 = 40/\gamma$, \footnotesize$\sigma_3 = 45/\gamma$, \footnotesize$\tau_1 = 400/\gamma$, \footnotesize$\tau_2 = 500/\gamma$, \footnotesize$\tau_3 = 500/\gamma$. \small Figures in a particular column have a common time axis.}
\end{center} 
\end{figure}

The propagation of a $\sech^2$, square and a complex shaped pulse, are investigated indiscriminately in presence of a continuous control field. The complex shaped pulse and $\sech^2$ pulse are given as: 

\begin{subequations}
\begin{align*}
\Omega_p(0,t) &= \Omega^0_p \left(e^{-(t-\tau_1)^2/\sigma_1^2} +  \sech[-(t-\tau_2)/\sigma_2]\right),\\
\Omega_p(0,t) &= \Omega^0_p \sech^2[-(t-\tau_3)/\sigma_3].
\end{align*}
\end{subequations}

In presence of a continuous control field, both the complex and $\sech^2$ pulses show similar amplification and elongation (Fig. \ref{fig:ancillary_results}) as obtained earlier with a Gaussian pulse. The control field also shows similar absorption as before. The square pulse shows similar elongation but doesn't retain it's square shape at end of amplification process. Therefore, our method can be used to amplify arbitrary pulse shapes without loss of generality.

Next, the propagation of a Gaussian pulse in presence of a continuous control field is investigated for different control magnitudes.
\begin{figure}[H]
\centering
\includegraphics[scale=1]{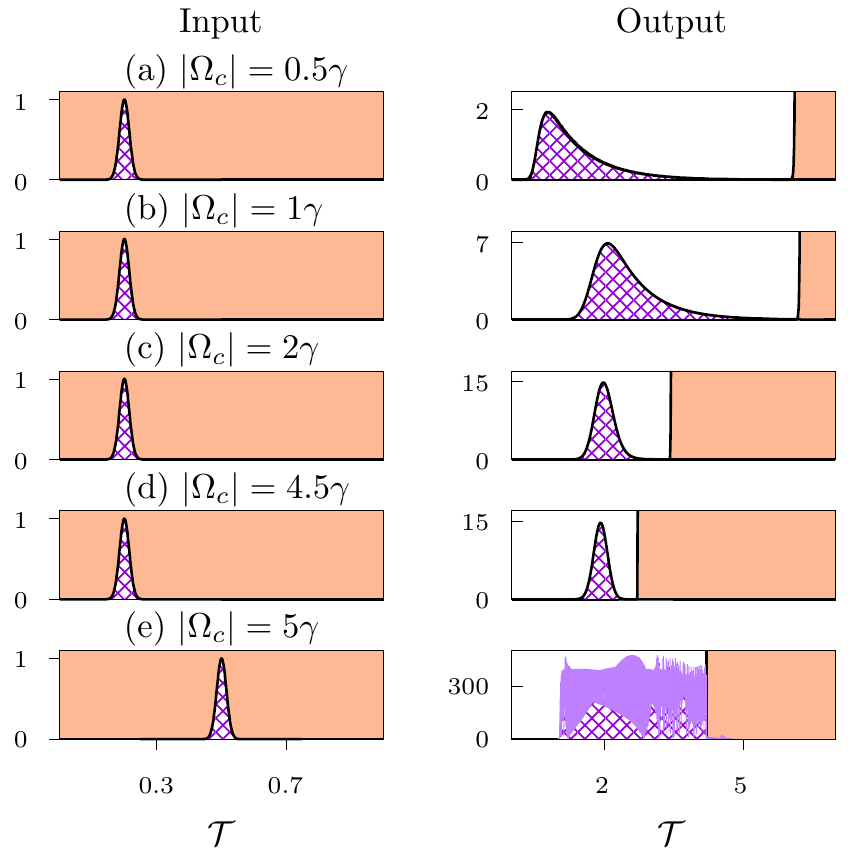}
\vspace*{0.1pt}
\caption{\label{fig:different_control_magnitude}(Color online) (left column) Temporal profiles of probe (crisscross) and control (solid) field magnitudes at $\mathcal{Z} = 0$, for different values of $|\Omega_c/\gamma|$. (right column) Corresponding temporal profiles of probe and control field magnitudes at the end of amplification process. Field magnitude scaling, normalization and relevant parameters remain same as Fig. \ref{fig:probe_amplification}. Figures in a particular column have a common time axis.}
\end{figure}

As explained earlier in Fig. \ref{fig:pulse_broadening_explaination}, the amplified portion of the probe (crisscross), being uninfluenced by the control, isn't subjected to pulse broadening due to dispersion but the portion of probe under the influence of control (oblique line pattern) continues to experience broadening. Therefore, if $\kappa_2$, which is responsible for pulse broadening is large, then the oblique line portion of probe will experience significant broadening, resulting in an asymmetric pulse shape posterior to the amplification process. This is shown in Fig. \ref{fig:different_control_magnitude}(a), where $\kappa_2 = 31.651 i$ for $|\Omega_c/\gamma| = 0.5$, results in an asymmetric output pulse shape. Whereas, the output pulse shape for $\kappa_2 = 4.89\times10^{-3}i$ corresponding to $|\Omega_c/\gamma| = 4.5$ in Fig. \ref{fig:different_control_magnitude}(d), shows closer semblance to the input Gaussian shape. The parameter $\kappa_2$ decreases with increase in $|\Omega_c|$. Therefore, in Fig. \ref{fig:different_control_magnitude}(a) - (d), as $|\Omega_c/\gamma|$ is increased from $0.5$ to $4.5$, the output pulse shape becomes more close to the input Gaussian shape.

In Fig. \ref{fig:different_control_magnitude}(a) - (e), as $|\Omega_c/\gamma|$ is increased from $0.5$ to $5$, the amplification also increases. Raising $|\Omega_c|$, beyond a certain value, e.g., $|\Omega_c|=5\gamma$ in Fig. \ref{fig:different_control_magnitude}(e), results in unstable output. The reason for this is explained in detail in appexdix \ref{app:B}.

Therefore, choice of $|\Omega_c|$ plays a crucial role in deciding the amount of amplification and the shape of probe pulse at the end of amplification process. A relatively small $|\Omega_c|$ leads to small amplification and asymmetricities in the output pulse shape. Again, a large $|\Omega_c|$ results in instabilities caused by huge gain. The sweet spot lies somewhere in the middle, as seen in Fig. \ref{fig:different_control_magnitude}.
%
%
%
\section{\label{sec:summary_and_conclusion}SUMMARY AND CONCLUSIONS}

In conclusion, the propagation of a weak probe pulse through the $\Lambda$ system in a resonant gain configuration is investigated. The gain configuration is different from the EIT system in a way that the control and probe fields are swapped with one another. This configuration makes provision for population inversion in the transition coupled by the probe field. Thus, causing probe amplification through stimulated emission. This broadening can be easily compensated using compressors. With a careful choice of control field intensity, the probe pulse although broadened, retains its initial shape and travels at the speed of light in vacuum, without any delay, absorption and dispersion after the amplification process. With this scheme, an arbitrary shaped probe pulse can propagate through the medium without loss of generality at the end of amplification. Therefore, the proposed model system could be useful in optical communication as one of the methods to amplify various signals, for possibly assisting long distance communication.
%
%

\begin{acknowledgments}
T.N.D. gratefully acknowledges funding by the Science and Engineering Board (Grant No. CRG/2018/000054).
\end{acknowledgments}

%
\appendix
\section{\label{app:A}Relation between $T$ and $\sigma$}
\begin{figure}[t]
\begin{center}
\includegraphics[scale=1]{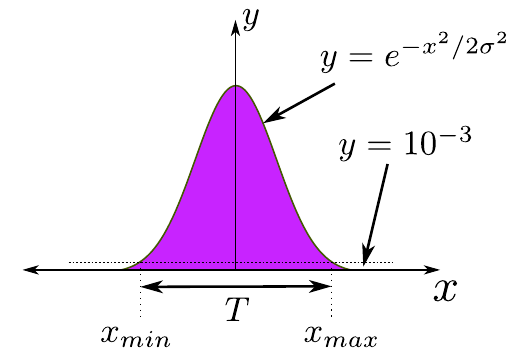}
\vspace*{2mm}
\caption{\label{fig:gaussian_distribution }Gaussian distribution function}
\end{center}
\end{figure}

In Fig. \ref{fig:gaussian_distribution }, the relation between the width $\sigma$ of a Gaussian function $(y = e^{-x^2/2\sigma^2})$ and the full extent of the Gaussian function $T$ can be approximated by considering an arbitrarily small value  say $(10^{-3})$, and then evaluating the values of $x_{min}$ and $x_{max}$, where  $x_{min}$ and $x_{max}$ are the $x$ coordinates of the intersection points of $y = 10^{-3}$ line with the Gaussian function.

\begin{figure}[b]
\includegraphics[scale=0.4]{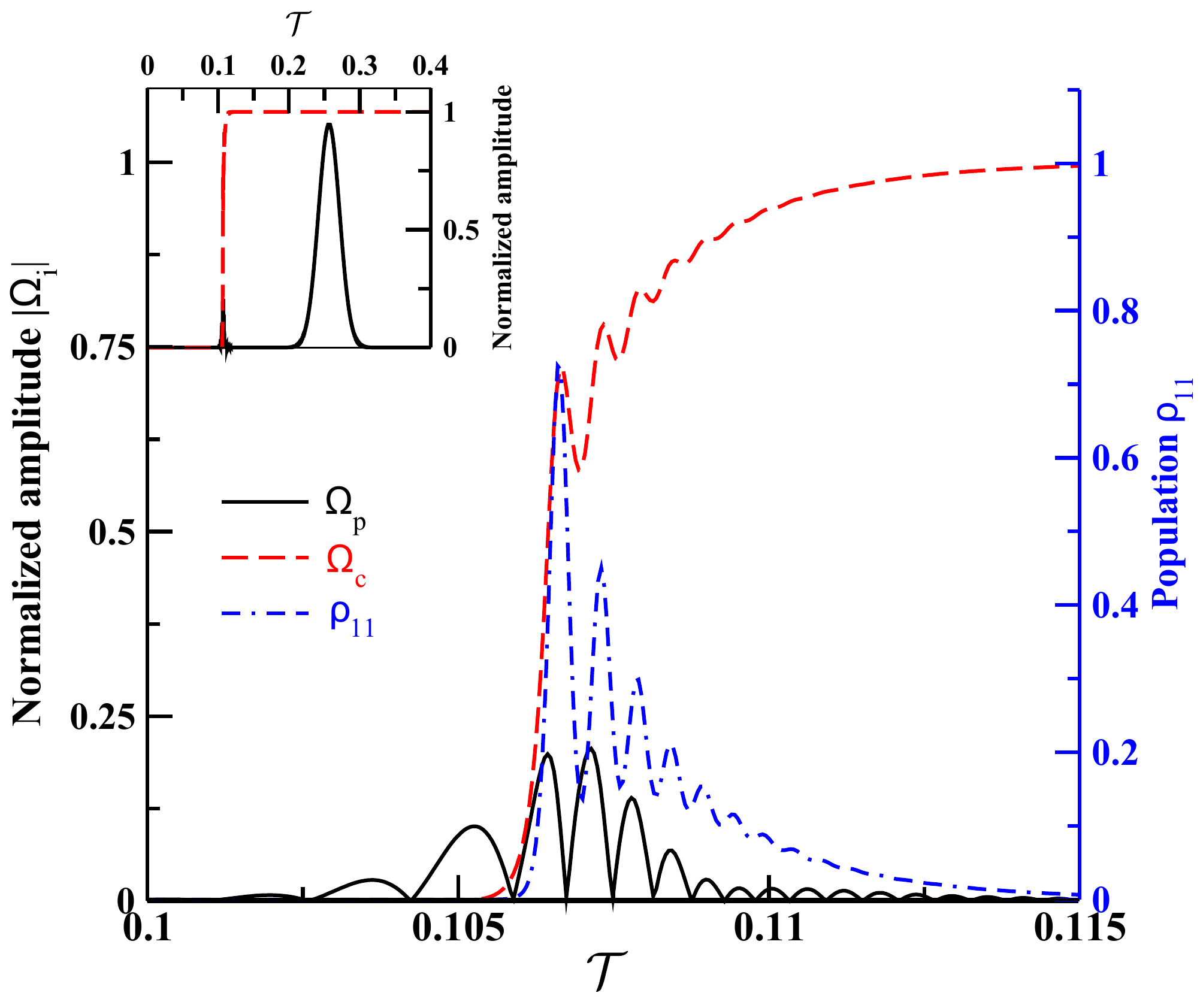}
\caption{\label{fig:early_amplification}(Color online) The normalized field amplitudes $(\Omega_i, i \in p,c)$ and population $\rho_{11}$ are plotted against normalized time $\mathcal T$ within the medium at $\mathcal{Z}= 2.8$. The inset shows temporal profiles of probe and control amplitudes in the vicinity of  $\mathcal T=0.1$.}
\end{figure}

\begin{figure}[b] 
\includegraphics[scale=0.45]{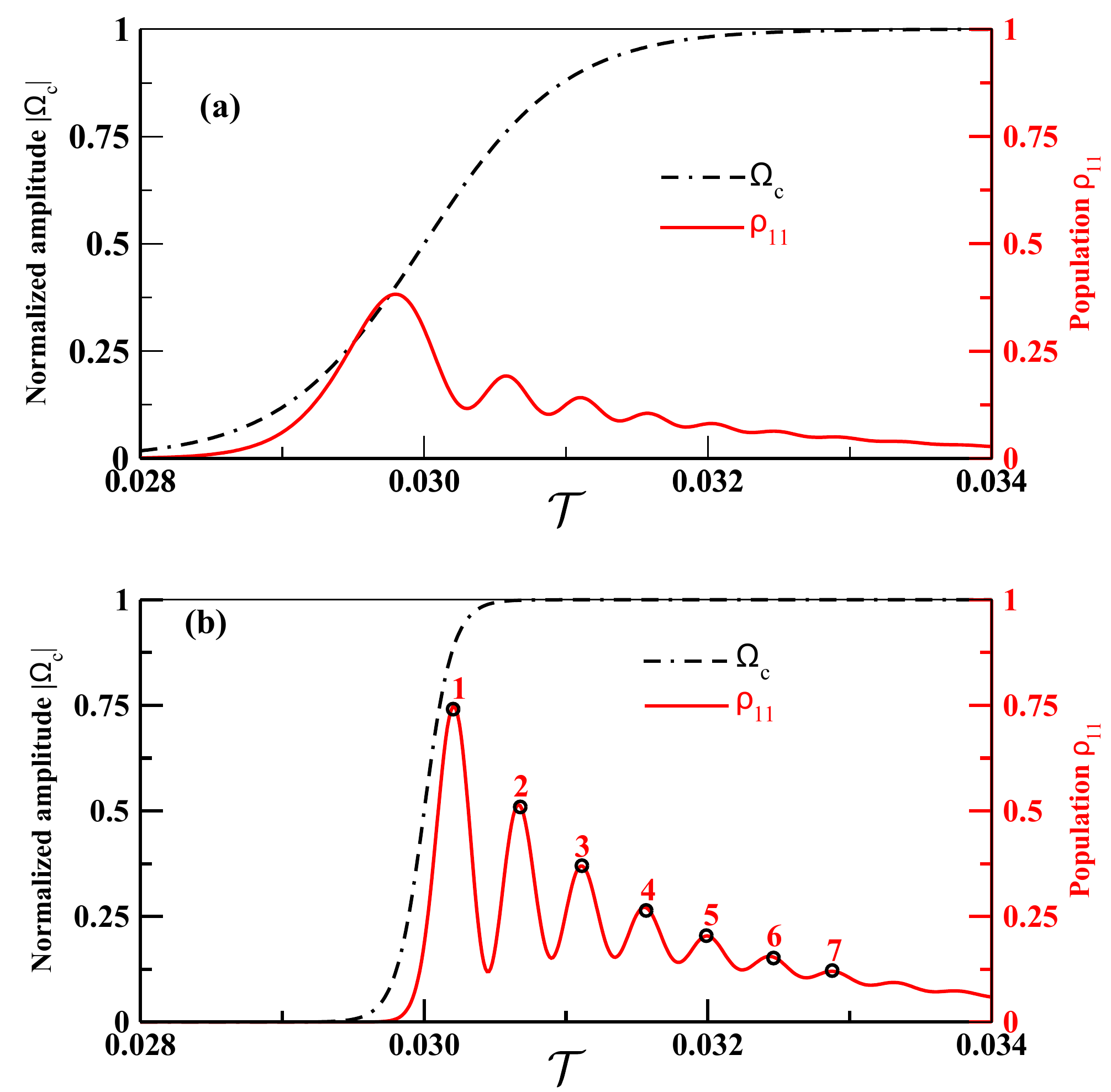}
\caption{\label{fig:adiabatic_transfer}(Color online) The normalized field amplitude $|\Omega_c|$ and population $\rho_{11}$ are plotted against normalized time $\mathcal T$ by numerically solving the density matrix equations for the system given in Fig. \ref{fig:level_diagram} with parameters $|\Omega_c| = 7\gamma$, $|\Omega_p| = 0.01\gamma$, and $\gamma_{23} = 0.001$. In Fig.\ref{fig:adiabatic_transfer}(a), $\alpha = 2$, representing a adiabatic switch whereas in Fig.\ref{fig:adiabatic_transfer}(b), $\alpha = 0.2 $ representing non adiabatic switch. The switching on time is $T_{on} = 30/\gamma$  }

\end{figure}

\noindent From Fig. \ref{fig:gaussian_distribution }
\begin{align}
&e^{-x^2_{min}/2\sigma^2} = 10^{-3}\notag\\
\implies & x_{min} = -\sqrt{2\sigma^2 \ln(10^3)},\label{eq:x_min}
\end{align}
and,
\begin{align}
&e^{-x^2_{max}/2\sigma^2} = 10^{-3}\notag\\
\implies &x_{max} = \sqrt{2\sigma^2 \ln(10^3)},\label{eq:x_max}
\end{align}

From Eqs. [(\ref{eq:x_min}), (\ref{eq:x_max})],
\begin{align}
&x_{max} - x_{min} = 2\sqrt{2\sigma^2 \ln(10^3)} \notag \\
\implies &T = 2\sqrt{2\ln(10^3)}\sigma.
\end{align}
%
\section{\label{app:B}Explanation for noisy output at high control field intensity}
\begin{figure}[b]
\includegraphics[scale=0.35]{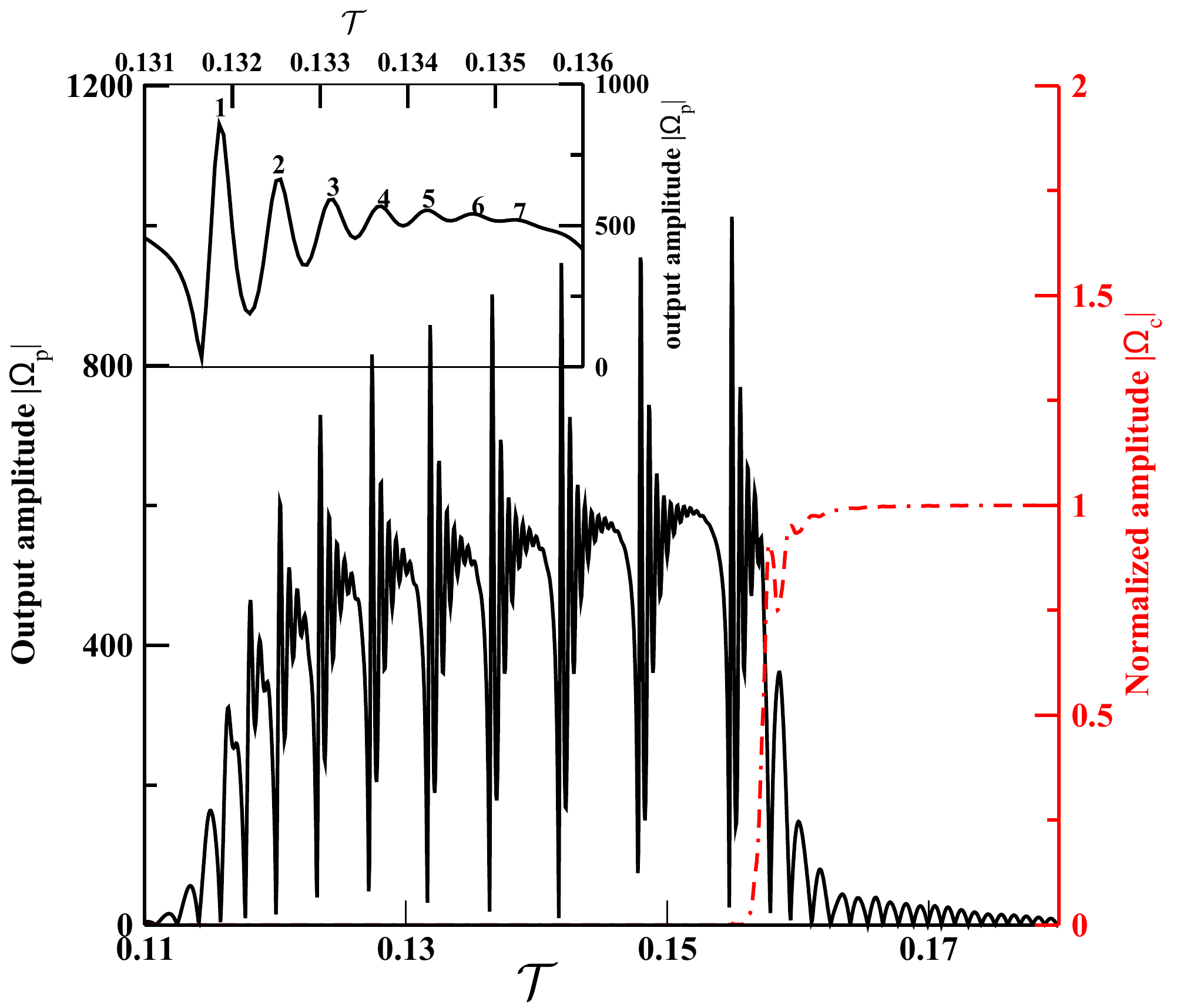}
\caption{\label{fig:noise}(Color online) Plot showing the output probe pulse for $\Omega^0_c = 7\gamma$ at $\mathcal{Z} = 4.5$. The probe is amplified around $1000$ times. Inset shows zoomed image of the probe field in the interval $0.131<\mathcal{T}<0.135$.}
\end{figure}

For control field amplitudes $5\gamma$ or higher, the probe pulse amplification begins a bit earlier as shown in the inset of Fig. \ref{fig:early_amplification}, where small traces of amplified probe pulse can be seen at around $\mathcal{T} = 0.1$. In Fig. \ref{fig:early_amplification}, at the position of control field's leading end on time axis, $\rho_{11}$ undergoes oscillations. These oscillations are due to the non adiabatic population transfer to the excited state $\ket{1}$ as a result of control field absorption. Such non adiabatic population transfer occurs when there is a steep rise in the control field intensity at its leading end instead of a smooth rise (adiabatic). Similar results have been reported by Harris \textit{et al.} \cite{preparation_energy} and Laine \textit{et al.} \cite{adiabatic_process}. In case of a smooth rise of control field intensity, such oscillations disappear as shown in Fig. \ref{fig:adiabatic_transfer}(a). Prominent  oscillations in $\rho_{11}$ is evident from Fig. \ref{fig:adiabatic_transfer}(b) for the non-adiabatic case. The adiabatic and non-adiabatic switching of the control field is possible by considering the envelope of the form $0.5\times\tanh((\tau - T_{on})/\alpha)$ whereas probe field envelope is taken to be continuous wave.  The parameter $\alpha$ determines how smoothly the control field is switched on. Larger the value of $\alpha$, smoother is the switching on. In Fig.\ref{fig:adiabatic_transfer}(a), $\alpha = 2$, causing an adiabatic transition and in Fig.\ref{fig:adiabatic_transfer}(b), $\alpha = 0.2 $ leading to a non adiabatic transition. Figure \ref{fig:adiabatic_transfer} is generated by merely solving the time dependent density matrix equations of the system given in Fig. \ref{fig:level_diagram}, without considering any propagation of the fields. In  Fig. \ref{fig:adiabatic_transfer}, it can be seen that the amplitude as well as frequency of population oscillation in the excited state $\ket{1}$ increases as the control field is switched on in a non adiabatic manner. In the problem at hand, a smooth rise of control field's amplitude is not possible because of inherent medium absorption during propagation. Its leading end eventually becomes steep, mimicking a non adiabatic rise in control field intensity, irrespective of how smoothly it is switched on at the entrance of the medium. Therefore, the oscillations in $\rho_{11}$ is inevitable for the current system. Due to the oscillations in $\rho_{11}$, the generated probe field also oscillates in a similar manner as $\rho_{11}$. The amplitude and frequency of population oscillation in $\ket{1}$ is less prounced, such that the small oscillations in the probe field doesn't receive noticeable amplification when the control field amplitude is optimum. But with a larger control field amplitude e.g. $\Omega^0_c = 7\gamma$, the amplification is large enough and the oscillations in the probe field gets amplified as shown in Fig. \ref{fig:noise}. Figure \ref{fig:noise} shows the temporal profiles of the probe and control fields at a larger propagation distance $\mathcal{Z}= 4.5$ where the probe receives significant amplification with noisy oscillations. Upon zooming (inset of Fig. \ref{fig:noise}), the probe can be seen oscillating in a similar manner as $\rho_{11}$ in Fig. \ref{fig:adiabatic_transfer}(b). Such behavior suggests that the modulation in probe amplitude is indeed due to the population oscillation in excited state $\ket{1}$.

	\bibliography{reference}

\begin{thebibliography}{41}%
\makeatletter
\providecommand \@ifxundefined [1]{%
 \@ifx{#1\undefined}
}%
\providecommand \@ifnum [1]{%
 \ifnum #1\expandafter \@firstoftwo
 \else \expandafter \@secondoftwo
 \fi
}%
\providecommand \@ifx [1]{%
 \ifx #1\expandafter \@firstoftwo
 \else \expandafter \@secondoftwo
 \fi
}%
\providecommand \natexlab [1]{#1}%
\providecommand \enquote  [1]{``#1''}%
\providecommand \bibnamefont  [1]{#1}%
\providecommand \bibfnamefont [1]{#1}%
\providecommand \citenamefont [1]{#1}%
\providecommand \href@noop [0]{\@secondoftwo}%
\providecommand \href [0]{\begingroup \@sanitize@url \@href}%
\providecommand \@href[1]{\@@startlink{#1}\@@href}%
\providecommand \@@href[1]{\endgroup#1\@@endlink}%
\providecommand \@sanitize@url [0]{\catcode `\\12\catcode `\$12\catcode
  `\&12\catcode `\#12\catcode `\^12\catcode `\_12\catcode `\%12\relax}%
\providecommand \@@startlink[1]{}%
\providecommand \@@endlink[0]{}%
\providecommand \url  [0]{\begingroup\@sanitize@url \@url }%
\providecommand \@url [1]{\endgroup\@href {#1}{\urlprefix }}%
\providecommand \urlprefix  [0]{URL }%
\providecommand \Eprint [0]{\href }%
\providecommand \doibase [0]{http://dx.doi.org/}%
\providecommand \selectlanguage [0]{\@gobble}%
\providecommand \bibinfo  [0]{\@secondoftwo}%
\providecommand \bibfield  [0]{\@secondoftwo}%
\providecommand \translation [1]{[#1]}%
\providecommand \BibitemOpen [0]{}%
\providecommand \bibitemStop [0]{}%
\providecommand \bibitemNoStop [0]{.\EOS\space}%
\providecommand \EOS [0]{\spacefactor3000\relax}%
\providecommand \BibitemShut  [1]{\csname bibitem#1\endcsname}%
\let\auto@bib@innerbib\@empty
\bibitem [{\citenamefont {Mok}\ and\ \citenamefont {Eggleton}(2005)}]{RN16567}%
  \BibitemOpen
  \bibfield  {author} {\bibinfo {author} {\bibfnamefont {J.~T.}\ \bibnamefont
  {Mok}}\ and\ \bibinfo {author} {\bibfnamefont {B.~J.}\ \bibnamefont
  {Eggleton}},\ }\href {\doibase 10.1038/433811a} {\bibfield  {journal}
  {\bibinfo  {journal} {Nature}\ }\textbf {\bibinfo {volume} {433}},\ \bibinfo
  {pages} {811} (\bibinfo {year} {2005})}\BibitemShut {NoStop}%
\bibitem [{\citenamefont {Boyd}(2009)}]{Boyd1}%
  \BibitemOpen
  \bibfield  {author} {\bibinfo {author} {\bibfnamefont {R.~W.}\ \bibnamefont
  {Boyd}},\ }\href {\doibase 10.1080/09500340903159495} {\bibfield  {journal}
  {\bibinfo  {journal} {Journal of Modern Optics}\ }\textbf {\bibinfo {volume}
  {56}},\ \bibinfo {pages} {1908} (\bibinfo {year} {2009})}\BibitemShut
  {NoStop}%
\bibitem [{\citenamefont {Manzoni}\ \emph {et~al.}(2015)\citenamefont
  {Manzoni}, \citenamefont {Mucke}, \citenamefont {Cirmi}, \citenamefont
  {Fang}, \citenamefont {Moses}, \citenamefont {Huang}, \citenamefont {Hong},
  \citenamefont {Cerullo},\ and\ \citenamefont {Kartner}}]{lpr1}%
  \BibitemOpen
  \bibfield  {author} {\bibinfo {author} {\bibfnamefont {C.}~\bibnamefont
  {Manzoni}}, \bibinfo {author} {\bibfnamefont {O.~D.}\ \bibnamefont {Mucke}},
  \bibinfo {author} {\bibfnamefont {G.}~\bibnamefont {Cirmi}}, \bibinfo
  {author} {\bibfnamefont {S.~B.}\ \bibnamefont {Fang}}, \bibinfo {author}
  {\bibfnamefont {J.}~\bibnamefont {Moses}}, \bibinfo {author} {\bibfnamefont
  {S.~W.}\ \bibnamefont {Huang}}, \bibinfo {author} {\bibfnamefont {K.~H.}\
  \bibnamefont {Hong}}, \bibinfo {author} {\bibfnamefont {G.}~\bibnamefont
  {Cerullo}}, \ and\ \bibinfo {author} {\bibfnamefont {F.~X.}\ \bibnamefont
  {Kartner}},\ }\href {\doibase https://doi.org/10.1002/lpor.201400181}
  {\bibfield  {journal} {\bibinfo  {journal} {Laser \& Photonics Reviews}\
  }\textbf {\bibinfo {volume} {9}},\ \bibinfo {pages} {129} (\bibinfo {year}
  {2015})}\BibitemShut {NoStop}%
\bibitem [{\citenamefont {Kocharovskaya}\ and\ \citenamefont
  {Mandel}(1990)}]{Olga_1990}%
  \BibitemOpen
  \bibfield  {author} {\bibinfo {author} {\bibfnamefont {O.}~\bibnamefont
  {Kocharovskaya}}\ and\ \bibinfo {author} {\bibfnamefont {P.}~\bibnamefont
  {Mandel}},\ }\href {\doibase 10.1103/PhysRevA.42.523} {\bibfield  {journal}
  {\bibinfo  {journal} {Phys. Rev. A}\ }\textbf {\bibinfo {volume} {42}},\
  \bibinfo {pages} {523} (\bibinfo {year} {1990})}\BibitemShut {NoStop}%
\bibitem [{\citenamefont {Fleischhauer}\ and\ \citenamefont
  {Lukin}(2000)}]{Lukin_2000}%
  \BibitemOpen
  \bibfield  {author} {\bibinfo {author} {\bibfnamefont {M.}~\bibnamefont
  {Fleischhauer}}\ and\ \bibinfo {author} {\bibfnamefont {M.~D.}\ \bibnamefont
  {Lukin}},\ }\href {\doibase 10.1103/PhysRevLett.84.5094} {\bibfield
  {journal} {\bibinfo  {journal} {Phys. Rev. Lett.}\ }\textbf {\bibinfo
  {volume} {84}},\ \bibinfo {pages} {5094} (\bibinfo {year}
  {2000})}\BibitemShut {NoStop}%
\bibitem [{\citenamefont {Vitanov}\ \emph {et~al.}(2017)\citenamefont
  {Vitanov}, \citenamefont {Rangelov}, \citenamefont {Shore},\ and\
  \citenamefont {Bergmann}}]{stirap_1}%
  \BibitemOpen
  \bibfield  {author} {\bibinfo {author} {\bibfnamefont {N.~V.}\ \bibnamefont
  {Vitanov}}, \bibinfo {author} {\bibfnamefont {A.~A.}\ \bibnamefont
  {Rangelov}}, \bibinfo {author} {\bibfnamefont {B.~W.}\ \bibnamefont {Shore}},
  \ and\ \bibinfo {author} {\bibfnamefont {K.}~\bibnamefont {Bergmann}},\
  }\href {\doibase 10.1103/RevModPhys.89.015006} {\bibfield  {journal}
  {\bibinfo  {journal} {Rev. Mod. Phys.}\ }\textbf {\bibinfo {volume} {89}},\
  \bibinfo {pages} {015006} (\bibinfo {year} {2017})}\BibitemShut {NoStop}%
\bibitem [{\citenamefont {Fleischhauer}\ \emph {et~al.}(2005)\citenamefont
  {Fleischhauer}, \citenamefont {Imamoglu},\ and\ \citenamefont
  {Marangos}}]{eit_1}%
  \BibitemOpen
  \bibfield  {author} {\bibinfo {author} {\bibfnamefont {M.}~\bibnamefont
  {Fleischhauer}}, \bibinfo {author} {\bibfnamefont {A.}~\bibnamefont
  {Imamoglu}}, \ and\ \bibinfo {author} {\bibfnamefont {J.~P.}\ \bibnamefont
  {Marangos}},\ }\href {\doibase 10.1103/RevModPhys.77.633} {\bibfield
  {journal} {\bibinfo  {journal} {Rev. Mod. Phys.}\ }\textbf {\bibinfo {volume}
  {77}},\ \bibinfo {pages} {633} (\bibinfo {year} {2005})}\BibitemShut
  {NoStop}%
\bibitem [{\citenamefont {Arimondo}(1996)}]{cpt1}%
  \BibitemOpen
  \bibfield  {author} {\bibinfo {author} {\bibfnamefont {E.}~\bibnamefont
  {Arimondo}},\ }\href {\doibase https://doi.org/10.1016/S0079-6638(08)70531-6}
  {\emph {\bibinfo {title} {V Coherent Population Trapping in Laser
  Spectroscopy}}},\ edited by\ \bibinfo {editor} {\bibfnamefont
  {E.}~\bibnamefont {Wolf}},\ \bibinfo {series} {Progress in Optics},
  Vol.~\bibinfo {volume} {35}\ (\bibinfo  {publisher} {Elsevier},\ \bibinfo
  {year} {1996})\ pp.\ \bibinfo {pages} {257 -- 354}\BibitemShut {NoStop}%
\bibitem [{\citenamefont {Agarwal}\ and\ \citenamefont {Dey}(2009)}]{sat1}%
  \BibitemOpen
  \bibfield  {author} {\bibinfo {author} {\bibfnamefont {G.~S.}\ \bibnamefont
  {Agarwal}}\ and\ \bibinfo {author} {\bibfnamefont {T.~N.}\ \bibnamefont
  {Dey}},\ }\href {\doibase https://doi.org/10.1002/lpor.200810041} {\bibfield
  {journal} {\bibinfo  {journal} {Laser \& Photonics Reviews}\ }\textbf
  {\bibinfo {volume} {3}},\ \bibinfo {pages} {287} (\bibinfo {year}
  {2009})}\BibitemShut {NoStop}%
\bibitem [{\citenamefont {Harris}\ and\ \citenamefont
  {Sokolov}(1998)}]{PhysRevLett.81.2894}%
  \BibitemOpen
  \bibfield  {author} {\bibinfo {author} {\bibfnamefont {S.~E.}\ \bibnamefont
  {Harris}}\ and\ \bibinfo {author} {\bibfnamefont {A.~V.}\ \bibnamefont
  {Sokolov}},\ }\href {\doibase 10.1103/PhysRevLett.81.2894} {\bibfield
  {journal} {\bibinfo  {journal} {Phys. Rev. Lett.}\ }\textbf {\bibinfo
  {volume} {81}},\ \bibinfo {pages} {2894} (\bibinfo {year}
  {1998})}\BibitemShut {NoStop}%
\bibitem [{\citenamefont {V.}\ and\ \citenamefont
  {Dey}(2016)}]{PhysRevA.94.053851}%
  \BibitemOpen
  \bibfield  {author} {\bibinfo {author} {\bibfnamefont {R.~K.}\ \bibnamefont
  {V.}}\ and\ \bibinfo {author} {\bibfnamefont {T.~N.}\ \bibnamefont {Dey}},\
  }\href {\doibase 10.1103/PhysRevA.94.053851} {\bibfield  {journal} {\bibinfo
  {journal} {Phys. Rev. A}\ }\textbf {\bibinfo {volume} {94}},\ \bibinfo
  {pages} {053851} (\bibinfo {year} {2016})}\BibitemShut {NoStop}%
\bibitem [{\citenamefont {{Murata}}\ \emph {et~al.}(2000)\citenamefont
  {{Murata}}, \citenamefont {{Morimoto}}, \citenamefont {{Kobayashi}},\ and\
  \citenamefont {{Yamamoto}}}]{pulse_generation1}%
  \BibitemOpen
  \bibfield  {author} {\bibinfo {author} {\bibfnamefont {H.}~\bibnamefont
  {{Murata}}}, \bibinfo {author} {\bibfnamefont {A.}~\bibnamefont
  {{Morimoto}}}, \bibinfo {author} {\bibfnamefont {T.}~\bibnamefont
  {{Kobayashi}}}, \ and\ \bibinfo {author} {\bibfnamefont {S.}~\bibnamefont
  {{Yamamoto}}},\ }\href {\doibase 10.1109/2944.902186} {\bibfield  {journal}
  {\bibinfo  {journal} {IEEE Journal of Selected Topics in Quantum
  Electronics}\ }\textbf {\bibinfo {volume} {6}},\ \bibinfo {pages} {1325}
  (\bibinfo {year} {2000})}\BibitemShut {NoStop}%
\bibitem [{\citenamefont {Harris}\ \emph {et~al.}(1993)\citenamefont {Harris},
  \citenamefont {Macklin},\ and\ \citenamefont {Hansch}}]{pulse-generation2}%
  \BibitemOpen
  \bibfield  {author} {\bibinfo {author} {\bibfnamefont {S.}~\bibnamefont
  {Harris}}, \bibinfo {author} {\bibfnamefont {J.}~\bibnamefont {Macklin}}, \
  and\ \bibinfo {author} {\bibfnamefont {T.}~\bibnamefont {Hansch}},\ }\href
  {\doibase https://doi.org/10.1016/0030-4018(93)90250-9} {\bibfield  {journal}
  {\bibinfo  {journal} {Optics Communications}\ }\textbf {\bibinfo {volume}
  {100}},\ \bibinfo {pages} {487 } (\bibinfo {year} {1993})}\BibitemShut
  {NoStop}%
\bibitem [{\citenamefont {McCall}\ and\ \citenamefont {Hahn}(1969)}]{McCall1}%
  \BibitemOpen
  \bibfield  {author} {\bibinfo {author} {\bibfnamefont {S.~L.}\ \bibnamefont
  {McCall}}\ and\ \bibinfo {author} {\bibfnamefont {E.~L.}\ \bibnamefont
  {Hahn}},\ }\href {\doibase 10.1103/PhysRev.183.457} {\bibfield  {journal}
  {\bibinfo  {journal} {Phys. Rev.}\ }\textbf {\bibinfo {volume} {183}},\
  \bibinfo {pages} {457} (\bibinfo {year} {1969})}\BibitemShut {NoStop}%
\bibitem [{\citenamefont {McCall}\ and\ \citenamefont {Hahn}(1967)}]{McCall2}%
  \BibitemOpen
  \bibfield  {author} {\bibinfo {author} {\bibfnamefont {S.~L.}\ \bibnamefont
  {McCall}}\ and\ \bibinfo {author} {\bibfnamefont {E.~L.}\ \bibnamefont
  {Hahn}},\ }\href {\doibase 10.1103/PhysRevLett.18.908} {\bibfield  {journal}
  {\bibinfo  {journal} {Phys. Rev. Lett.}\ }\textbf {\bibinfo {volume} {18}},\
  \bibinfo {pages} {908} (\bibinfo {year} {1967})}\BibitemShut {NoStop}%
\bibitem [{\citenamefont {Konopnicki}\ and\ \citenamefont
  {Eberly}(1981)}]{PhysRevA.24.2567}%
  \BibitemOpen
  \bibfield  {author} {\bibinfo {author} {\bibfnamefont {M.~J.}\ \bibnamefont
  {Konopnicki}}\ and\ \bibinfo {author} {\bibfnamefont {J.~H.}\ \bibnamefont
  {Eberly}},\ }\href {\doibase 10.1103/PhysRevA.24.2567} {\bibfield  {journal}
  {\bibinfo  {journal} {Phys. Rev. A}\ }\textbf {\bibinfo {volume} {24}},\
  \bibinfo {pages} {2567} (\bibinfo {year} {1981})}\BibitemShut {NoStop}%
\bibitem [{\citenamefont {Grobe}\ \emph {et~al.}(1994)\citenamefont {Grobe},
  \citenamefont {Hioe},\ and\ \citenamefont {Eberly}}]{PhysRevLett.73.3183}%
  \BibitemOpen
  \bibfield  {author} {\bibinfo {author} {\bibfnamefont {R.}~\bibnamefont
  {Grobe}}, \bibinfo {author} {\bibfnamefont {F.~T.}\ \bibnamefont {Hioe}}, \
  and\ \bibinfo {author} {\bibfnamefont {J.~H.}\ \bibnamefont {Eberly}},\
  }\href {\doibase 10.1103/PhysRevLett.73.3183} {\bibfield  {journal} {\bibinfo
   {journal} {Phys. Rev. Lett.}\ }\textbf {\bibinfo {volume} {73}},\ \bibinfo
  {pages} {3183} (\bibinfo {year} {1994})}\BibitemShut {NoStop}%
\bibitem [{\citenamefont {Eberly}(1995)}]{Eberly_1995}%
  \BibitemOpen
  \bibfield  {author} {\bibinfo {author} {\bibfnamefont {J.~H.}\ \bibnamefont
  {Eberly}},\ }\href {\doibase 10.1088/1355-5111/7/3/013} {\bibfield  {journal}
  {\bibinfo  {journal} {Quantum and Semiclassical Optics: Journal of the
  European Optical Society Part B}\ }\textbf {\bibinfo {volume} {7}},\ \bibinfo
  {pages} {373} (\bibinfo {year} {1995})}\BibitemShut {NoStop}%
\bibitem [{\citenamefont {Rahman}\ and\ \citenamefont
  {Eberly}(1998)}]{PhysRevA.58.R805}%
  \BibitemOpen
  \bibfield  {author} {\bibinfo {author} {\bibfnamefont {A.}~\bibnamefont
  {Rahman}}\ and\ \bibinfo {author} {\bibfnamefont {J.~H.}\ \bibnamefont
  {Eberly}},\ }\href {\doibase 10.1103/PhysRevA.58.R805} {\bibfield  {journal}
  {\bibinfo  {journal} {Phys. Rev. A}\ }\textbf {\bibinfo {volume} {58}},\
  \bibinfo {pages} {R805} (\bibinfo {year} {1998})}\BibitemShut {NoStop}%
\bibitem [{\citenamefont {Park}\ and\ \citenamefont
  {Shin}(1998)}]{PhysRevA.57.4643}%
  \BibitemOpen
  \bibfield  {author} {\bibinfo {author} {\bibfnamefont {Q.-H.}\ \bibnamefont
  {Park}}\ and\ \bibinfo {author} {\bibfnamefont {H.~J.}\ \bibnamefont
  {Shin}},\ }\href {\doibase 10.1103/PhysRevA.57.4643} {\bibfield  {journal}
  {\bibinfo  {journal} {Phys. Rev. A}\ }\textbf {\bibinfo {volume} {57}},\
  \bibinfo {pages} {4643} (\bibinfo {year} {1998})}\BibitemShut {NoStop}%
\bibitem [{\citenamefont {Rahman}(1999)}]{PhysRevA.60.4187}%
  \BibitemOpen
  \bibfield  {author} {\bibinfo {author} {\bibfnamefont {A.}~\bibnamefont
  {Rahman}},\ }\href {\doibase 10.1103/PhysRevA.60.4187} {\bibfield  {journal}
  {\bibinfo  {journal} {Phys. Rev. A}\ }\textbf {\bibinfo {volume} {60}},\
  \bibinfo {pages} {4187} (\bibinfo {year} {1999})}\BibitemShut {NoStop}%
\bibitem [{\citenamefont {Agarwal}\ and\ \citenamefont
  {Eberly}(1999)}]{agarwal1}%
  \BibitemOpen
  \bibfield  {author} {\bibinfo {author} {\bibfnamefont {G.~S.}\ \bibnamefont
  {Agarwal}}\ and\ \bibinfo {author} {\bibfnamefont {J.~H.}\ \bibnamefont
  {Eberly}},\ }\href {\doibase 10.1103/PhysRevA.61.013404} {\bibfield
  {journal} {\bibinfo  {journal} {Phys. Rev. A}\ }\textbf {\bibinfo {volume}
  {61}},\ \bibinfo {pages} {013404} (\bibinfo {year} {1999})}\BibitemShut
  {NoStop}%
\bibitem [{\citenamefont {Vemuri}\ \emph {et~al.}(1997)\citenamefont {Vemuri},
  \citenamefont {Agarwal},\ and\ \citenamefont {Vasavada}}]{cloning1}%
  \BibitemOpen
  \bibfield  {author} {\bibinfo {author} {\bibfnamefont {G.}~\bibnamefont
  {Vemuri}}, \bibinfo {author} {\bibfnamefont {G.~S.}\ \bibnamefont {Agarwal}},
  \ and\ \bibinfo {author} {\bibfnamefont {K.~V.}\ \bibnamefont {Vasavada}},\
  }\href {\doibase 10.1103/PhysRevLett.79.3889} {\bibfield  {journal} {\bibinfo
   {journal} {Phys. Rev. Lett.}\ }\textbf {\bibinfo {volume} {79}},\ \bibinfo
  {pages} {3889} (\bibinfo {year} {1997})}\BibitemShut {NoStop}%
\bibitem [{\citenamefont {Ogden}\ \emph {et~al.}(2019)\citenamefont {Ogden},
  \citenamefont {Whittaker}, \citenamefont {Keaveney}, \citenamefont
  {Wrathmall}, \citenamefont {Adams},\ and\ \citenamefont
  {Potvliege}}]{simultons2}%
  \BibitemOpen
  \bibfield  {author} {\bibinfo {author} {\bibfnamefont {T.~P.}\ \bibnamefont
  {Ogden}}, \bibinfo {author} {\bibfnamefont {K.~A.}\ \bibnamefont
  {Whittaker}}, \bibinfo {author} {\bibfnamefont {J.}~\bibnamefont {Keaveney}},
  \bibinfo {author} {\bibfnamefont {S.~A.}\ \bibnamefont {Wrathmall}}, \bibinfo
  {author} {\bibfnamefont {C.~S.}\ \bibnamefont {Adams}}, \ and\ \bibinfo
  {author} {\bibfnamefont {R.~M.}\ \bibnamefont {Potvliege}},\ }\href {\doibase
  10.1103/PhysRevLett.123.243604} {\bibfield  {journal} {\bibinfo  {journal}
  {Phys. Rev. Lett.}\ }\textbf {\bibinfo {volume} {123}},\ \bibinfo {pages}
  {243604} (\bibinfo {year} {2019})}\BibitemShut {NoStop}%
\bibitem [{\citenamefont {Matsko}\ \emph {et~al.}(2001)\citenamefont {Matsko},
  \citenamefont {Kocharovskaya}, \citenamefont {Rostovtsev}, \citenamefont
  {Welch}, \citenamefont {Zibrov},\ and\ \citenamefont
  {Scully}}]{MATSKO2001191}%
  \BibitemOpen
  \bibfield  {author} {\bibinfo {author} {\bibfnamefont {A.~B.}\ \bibnamefont
  {Matsko}}, \bibinfo {author} {\bibfnamefont {O.}~\bibnamefont
  {Kocharovskaya}}, \bibinfo {author} {\bibfnamefont {Y.}~\bibnamefont
  {Rostovtsev}}, \bibinfo {author} {\bibfnamefont {G.~R.}\ \bibnamefont
  {Welch}}, \bibinfo {author} {\bibfnamefont {A.~S.}\ \bibnamefont {Zibrov}}, \
  and\ \bibinfo {author} {\bibfnamefont {M.~O.}\ \bibnamefont {Scully}},\
  }\href {\doibase https://doi.org/10.1016/S1049-250X(01)80064-1} {\emph
  {\bibinfo {title} {Slow, Ultraslow, Stored, and Frozen Light}}},\ edited by\
  \bibinfo {editor} {\bibfnamefont {B.}~\bibnamefont {Bederson}}\ and\ \bibinfo
  {editor} {\bibfnamefont {H.}~\bibnamefont {Walther}},\ \bibinfo {series}
  {Advances In Atomic, Molecular, and Optical Physics}, Vol.~\bibinfo {volume}
  {46}\ (\bibinfo  {publisher} {Academic Press},\ \bibinfo {year} {2001})\ pp.\
  \bibinfo {pages} {191 -- 242}\BibitemShut {NoStop}%
\bibitem [{\citenamefont {Harris}(1993)}]{matched1}%
  \BibitemOpen
  \bibfield  {author} {\bibinfo {author} {\bibfnamefont {S.~E.}\ \bibnamefont
  {Harris}},\ }\href {\doibase 10.1103/PhysRevLett.70.552} {\bibfield
  {journal} {\bibinfo  {journal} {Phys. Rev. Lett.}\ }\textbf {\bibinfo
  {volume} {70}},\ \bibinfo {pages} {552} (\bibinfo {year} {1993})}\BibitemShut
  {NoStop}%
\bibitem [{\citenamefont {Sun}\ \emph {et~al.}(2011)\citenamefont {Sun},
  \citenamefont {Sariyanni}, \citenamefont {Das},\ and\ \citenamefont
  {Rostovtsev}}]{Rostovtsev_2011}%
  \BibitemOpen
  \bibfield  {author} {\bibinfo {author} {\bibfnamefont {D.}~\bibnamefont
  {Sun}}, \bibinfo {author} {\bibfnamefont {Z.-E.}\ \bibnamefont {Sariyanni}},
  \bibinfo {author} {\bibfnamefont {S.}~\bibnamefont {Das}}, \ and\ \bibinfo
  {author} {\bibfnamefont {Y.~V.}\ \bibnamefont {Rostovtsev}},\ }\href
  {\doibase 10.1103/PhysRevA.83.063815} {\bibfield  {journal} {\bibinfo
  {journal} {Phys. Rev. A}\ }\textbf {\bibinfo {volume} {83}},\ \bibinfo
  {pages} {063815} (\bibinfo {year} {2011})}\BibitemShut {NoStop}%
\bibitem [{\citenamefont {Tanaka}\ \emph {et~al.}(2003)\citenamefont {Tanaka},
  \citenamefont {Niwa}, \citenamefont {Hayami}, \citenamefont {Furue},
  \citenamefont {Nakayama}, \citenamefont {Kohmoto}, \citenamefont {Kunitomo},\
  and\ \citenamefont {Fukuda}}]{Tanaka_2003}%
  \BibitemOpen
  \bibfield  {author} {\bibinfo {author} {\bibfnamefont {H.}~\bibnamefont
  {Tanaka}}, \bibinfo {author} {\bibfnamefont {H.}~\bibnamefont {Niwa}},
  \bibinfo {author} {\bibfnamefont {K.}~\bibnamefont {Hayami}}, \bibinfo
  {author} {\bibfnamefont {S.}~\bibnamefont {Furue}}, \bibinfo {author}
  {\bibfnamefont {K.}~\bibnamefont {Nakayama}}, \bibinfo {author}
  {\bibfnamefont {T.}~\bibnamefont {Kohmoto}}, \bibinfo {author} {\bibfnamefont
  {M.}~\bibnamefont {Kunitomo}}, \ and\ \bibinfo {author} {\bibfnamefont
  {Y.}~\bibnamefont {Fukuda}},\ }\href {\doibase 10.1103/PhysRevA.68.053801}
  {\bibfield  {journal} {\bibinfo  {journal} {Phys. Rev. A}\ }\textbf {\bibinfo
  {volume} {68}},\ \bibinfo {pages} {053801} (\bibinfo {year}
  {2003})}\BibitemShut {NoStop}%
\bibitem [{\citenamefont {Wang}\ \emph {et~al.}(2000)\citenamefont {Wang},
  \citenamefont {Kuzmich},\ and\ \citenamefont {Dogariu}}]{Wang_2000}%
  \BibitemOpen
  \bibfield  {author} {\bibinfo {author} {\bibfnamefont {L.~J.}\ \bibnamefont
  {Wang}}, \bibinfo {author} {\bibfnamefont {A.}~\bibnamefont {Kuzmich}}, \
  and\ \bibinfo {author} {\bibfnamefont {A.}~\bibnamefont {Dogariu}},\ }\href
  {\doibase 10.1038/35018520} {\bibfield  {journal} {\bibinfo  {journal}
  {Nature}\ }\textbf {\bibinfo {volume} {406}},\ \bibinfo {pages} {277}
  (\bibinfo {year} {2000})}\BibitemShut {NoStop}%
\bibitem [{\citenamefont {Agarwal}\ \emph {et~al.}(2001)\citenamefont
  {Agarwal}, \citenamefont {Dey},\ and\ \citenamefont
  {Menon}}]{PhysRevA.64.053809}%
  \BibitemOpen
  \bibfield  {author} {\bibinfo {author} {\bibfnamefont {G.~S.}\ \bibnamefont
  {Agarwal}}, \bibinfo {author} {\bibfnamefont {T.~N.}\ \bibnamefont {Dey}}, \
  and\ \bibinfo {author} {\bibfnamefont {S.}~\bibnamefont {Menon}},\ }\href
  {\doibase 10.1103/PhysRevA.64.053809} {\bibfield  {journal} {\bibinfo
  {journal} {Phys. Rev. A}\ }\textbf {\bibinfo {volume} {64}},\ \bibinfo
  {pages} {053809} (\bibinfo {year} {2001})}\BibitemShut {NoStop}%
\bibitem [{\citenamefont {Clader}\ and\ \citenamefont
  {Eberly}(2008)}]{Eberly_2008}%
  \BibitemOpen
  \bibfield  {author} {\bibinfo {author} {\bibfnamefont {B.~D.}\ \bibnamefont
  {Clader}}\ and\ \bibinfo {author} {\bibfnamefont {J.~H.}\ \bibnamefont
  {Eberly}},\ }\href {\doibase 10.1103/PhysRevA.78.033803} {\bibfield
  {journal} {\bibinfo  {journal} {Phys. Rev. A}\ }\textbf {\bibinfo {volume}
  {78}},\ \bibinfo {pages} {033803} (\bibinfo {year} {2008})}\BibitemShut
  {NoStop}%
\bibitem [{\citenamefont {Eilam}\ and\ \citenamefont
  {Wilson-Gordon}(2018)}]{Gordon_2018}%
  \BibitemOpen
  \bibfield  {author} {\bibinfo {author} {\bibfnamefont {A.}~\bibnamefont
  {Eilam}}\ and\ \bibinfo {author} {\bibfnamefont {A.~D.}\ \bibnamefont
  {Wilson-Gordon}},\ }\href {\doibase 10.1103/PhysRevA.98.013808} {\bibfield
  {journal} {\bibinfo  {journal} {Phys. Rev. A}\ }\textbf {\bibinfo {volume}
  {98}},\ \bibinfo {pages} {013808} (\bibinfo {year} {2018})}\BibitemShut
  {NoStop}%
\bibitem [{\citenamefont {Hamedi}\ \emph {et~al.}(2017)\citenamefont {Hamedi},
  \citenamefont {Ruseckas},\ and\ \citenamefont
  {Juzeli{\={u}}nas}}]{Hamedi_2017}%
  \BibitemOpen
  \bibfield  {author} {\bibinfo {author} {\bibfnamefont {H.~R.}\ \bibnamefont
  {Hamedi}}, \bibinfo {author} {\bibfnamefont {J.}~\bibnamefont {Ruseckas}}, \
  and\ \bibinfo {author} {\bibfnamefont {G.}~\bibnamefont {Juzeli{\={u}}nas}},\
  }\href {\doibase 10.1088/1361-6455/aa84f6} {\bibfield  {journal} {\bibinfo
  {journal} {Journal of Physics B: Atomic, Molecular and Optical Physics}\
  }\textbf {\bibinfo {volume} {50}},\ \bibinfo {pages} {185401} (\bibinfo
  {year} {2017})}\BibitemShut {NoStop}%
\bibitem [{\citenamefont {Tai}\ \emph {et~al.}(1986)\citenamefont {Tai},
  \citenamefont {Hasegawa},\ and\ \citenamefont {Tomita}}]{instability1}%
  \BibitemOpen
  \bibfield  {author} {\bibinfo {author} {\bibfnamefont {K.}~\bibnamefont
  {Tai}}, \bibinfo {author} {\bibfnamefont {A.}~\bibnamefont {Hasegawa}}, \
  and\ \bibinfo {author} {\bibfnamefont {A.}~\bibnamefont {Tomita}},\ }\href
  {\doibase 10.1103/PhysRevLett.56.135} {\bibfield  {journal} {\bibinfo
  {journal} {Phys. Rev. Lett.}\ }\textbf {\bibinfo {volume} {56}},\ \bibinfo
  {pages} {135} (\bibinfo {year} {1986})}\BibitemShut {NoStop}%
\bibitem [{\citenamefont {Agrawal}(1987)}]{instability2}%
  \BibitemOpen
  \bibfield  {author} {\bibinfo {author} {\bibfnamefont {G.~P.}\ \bibnamefont
  {Agrawal}},\ }\href {\doibase 10.1103/PhysRevLett.59.880} {\bibfield
  {journal} {\bibinfo  {journal} {Phys. Rev. Lett.}\ }\textbf {\bibinfo
  {volume} {59}},\ \bibinfo {pages} {880} (\bibinfo {year} {1987})}\BibitemShut
  {NoStop}%
\bibitem [{\citenamefont {Boyd}(2008)}]{boyd_chapter_7}%
  \BibitemOpen
  \bibfield  {author} {\bibinfo {author} {\bibfnamefont {R.~W.}\ \bibnamefont
  {Boyd}},\ }\href {\doibase
  https://doi.org/10.1016/B978-0-12-369470-6.00007-1} {\emph {\bibinfo {title}
  {Nonlinear Optics (3\textsuperscript{rd} Edition)}}},\ \bibinfo {edition}
  {3rd}\ ed.,\ edited by\ \bibinfo {editor} {\bibfnamefont {R.~W.}\
  \bibnamefont {Boyd}}\ (\bibinfo  {publisher} {Academic Press},\ \bibinfo
  {address} {Burlington},\ \bibinfo {year} {2008})\ pp.\ \bibinfo {pages}
  {329--390}\BibitemShut {NoStop}%
\bibitem [{\citenamefont {Jiang}\ \emph {et~al.}(2006)\citenamefont {Jiang},
  \citenamefont {Deng},\ and\ \citenamefont {Payne}}]{arg_1}%
  \BibitemOpen
  \bibfield  {author} {\bibinfo {author} {\bibfnamefont {K.~J.}\ \bibnamefont
  {Jiang}}, \bibinfo {author} {\bibfnamefont {L.}~\bibnamefont {Deng}}, \ and\
  \bibinfo {author} {\bibfnamefont {M.~G.}\ \bibnamefont {Payne}},\ }\href
  {\doibase 10.1103/PhysRevA.74.041803} {\bibfield  {journal} {\bibinfo
  {journal} {Phys. Rev. A}\ }\textbf {\bibinfo {volume} {74}},\ \bibinfo
  {pages} {041803(R)} (\bibinfo {year} {2006})}\BibitemShut {NoStop}%
\bibitem [{\citenamefont {Li}\ \emph {et~al.}(2008)\citenamefont {Li},
  \citenamefont {Hang}, \citenamefont {Huang},\ and\ \citenamefont
  {Deng}}]{arg_3}%
  \BibitemOpen
  \bibfield  {author} {\bibinfo {author} {\bibfnamefont {H.-j.}\ \bibnamefont
  {Li}}, \bibinfo {author} {\bibfnamefont {C.}~\bibnamefont {Hang}}, \bibinfo
  {author} {\bibfnamefont {G.}~\bibnamefont {Huang}}, \ and\ \bibinfo {author}
  {\bibfnamefont {L.}~\bibnamefont {Deng}},\ }\href {\doibase
  10.1103/PhysRevA.78.023822} {\bibfield  {journal} {\bibinfo  {journal} {Phys.
  Rev. A}\ }\textbf {\bibinfo {volume} {78}},\ \bibinfo {pages} {023822}
  (\bibinfo {year} {2008})}\BibitemShut {NoStop}%
\bibitem [{\citenamefont {Strickland}\ and\ \citenamefont
  {Mourou}(1985)}]{Mourou_et_al}%
  \BibitemOpen
  \bibfield  {author} {\bibinfo {author} {\bibfnamefont {D.}~\bibnamefont
  {Strickland}}\ and\ \bibinfo {author} {\bibfnamefont {G.}~\bibnamefont
  {Mourou}},\ }\href {\doibase https://doi.org/10.1016/0030-4018(85)90120-8}
  {\bibfield  {journal} {\bibinfo  {journal} {Optics Communications}\ }\textbf
  {\bibinfo {volume} {56}},\ \bibinfo {pages} {219} (\bibinfo {year}
  {1985})}\BibitemShut {NoStop}%
\bibitem [{\citenamefont {Harris}\ and\ \citenamefont
  {Luo}(1995)}]{preparation_energy}%
  \BibitemOpen
  \bibfield  {author} {\bibinfo {author} {\bibfnamefont {S.~E.}\ \bibnamefont
  {Harris}}\ and\ \bibinfo {author} {\bibfnamefont {Z.-F.}\ \bibnamefont
  {Luo}},\ }\href {\doibase 10.1103/PhysRevA.52.R928} {\bibfield  {journal}
  {\bibinfo  {journal} {Phys. Rev. A}\ }\textbf {\bibinfo {volume} {52}},\
  \bibinfo {pages} {R928} (\bibinfo {year} {1995})}\BibitemShut {NoStop}%
\bibitem [{\citenamefont {Laine}\ and\ \citenamefont
  {Stenholm}(1996)}]{adiabatic_process}%
  \BibitemOpen
  \bibfield  {author} {\bibinfo {author} {\bibfnamefont {T.~A.}\ \bibnamefont
  {Laine}}\ and\ \bibinfo {author} {\bibfnamefont {S.}~\bibnamefont
  {Stenholm}},\ }\href {\doibase 10.1103/PhysRevA.53.2501} {\bibfield
  {journal} {\bibinfo  {journal} {Phys. Rev. A}\ }\textbf {\bibinfo {volume}
  {53}},\ \bibinfo {pages} {2501} (\bibinfo {year} {1996})}\BibitemShut
  {NoStop}%
\end{thebibliography}%

\end{document}